\documentclass[preprint,aps,showpacs]{revtex4}
\usepackage{mathrsfs}
\usepackage{amsmath}
\usepackage{amssymb}
\usepackage{epsfig}
\usepackage{graphicx}
\usepackage{booktabs}
\usepackage{array}
\usepackage{multirow}
\begin{document}
\renewcommand{\arraystretch}{0.5}
\newcommand{\beq}{\begin{eqnarray}}
\newcommand{\eeq}{\end{eqnarray}}
\newcommand{\non}{\nonumber\\ }

\newcommand{\acp}{ {\cal A}_{CP} }
\newcommand{\psl}{ p \hspace{-1.8truemm}/ }
\newcommand{\nsl}{ n \hspace{-2.2truemm}/ }
\newcommand{\vsl}{ v \hspace{-2.2truemm}/ }
\newcommand{\epsl}{\epsilon \hspace{-1.8truemm}/\,  }

\def \cpl{ Chin. Phys. Lett.  }
\def \ctp{ Commun. Theor. Phys.  }
\def \epjc{ Eur. Phys. J. C }
\def \jpg{  J. Phys. G }
\def \npb{  Nucl. Phys. B }
\def \plb{  Phys. Lett. B }
\def \prd{  Phys. Rev. D }
\def \prl{  Phys. Rev. Lett.  }
\def \zpc{  Z. Phys. C }
\def \jhep{ J. High Energy Phys.  }

\title{The $B(B_{s})\rightarrow D_{(s)}(\bar{D}_{(s)})\,T$ and $D_{(s)}^{\ast}(\bar{D}_{(s)}^{*})\,T$ decays in perturbative QCD approach}
\author{Zhi-Tian Zou}
\author{Xin Yu}
\author{Cai-Dian L\"{u}}\email{lucd@ihep.ac.cn}
\affiliation{Institute  of  High  Energy  Physics  and  Theoretical  Physics Center for Science Facilities,
Chinese Academy of Sciences, Beijing 100049, People's Republic of China }

\date{\today}
\begin{abstract}
In perturbative QCD approach, we investigate the  $B(B_{s})\rightarrow
D_{(s)}(\bar{D}_{(s)})\,T$ and
$D_{(s)}^{\ast}(\bar{D}_{(s)}^{*})\,T$ decays, which include the Cabibbo-Kobayashi-Maskawa (CKM) favored
decays and the Cabibbo-Kobayashi-Maskawa-suppressed decays, where T
denotes a light tensor meson. From our calculation, we find that the
nonfactorizable emission diagrams and the annihilation type diagrams
are important, especially for those color suppressed channels. For
those decays with a tensor meson emitted, the factorizable emission
diagrams vanish owing to the fact that a tensor meson can not be
produced through the local (V-A) or tensor current. The numerical
results show that the predictions for the branching ratios of
considered charmed B decays are in the range of $10^{-4}$ to
$10^{-6}$ for those CKM-favored decays (governed by $|V_{cb}|$) and
in the range of $10^{-5}$ to $10^{-8}$ for those CKM-suppressed
decays (governed by $|V_{ub}|$). We also predict large transverse
polarization contributions in many of the $B(B_{s}) \rightarrow
D_{(s)}^{\ast}(\bar{D}_{(s)}^{*})T$ decay channels.
\end{abstract}



\maketitle

\section{Introduction}

In the recent years, several experimental measurements about B decay
modes
 involving a light tensor meson (T) have been
 obtained \cite{jpg37075021}.
  These light tensor mesons include the isovector $a_{2}(1320)$, the isodoublet
 $K_{2}^{*}(1430)$, and isosinglet $f_{2}(1270)$ and
 $f_{2}^{\prime}(1525)$ \cite{jpg37075021}. For the tensor meson with $J^{p}=2^{+}$, both the orbital
 angular momentum L and the total spin S of the quark pair are equal
 to 1. However, their production property in B decays is
 quite similar to the light vector mesons \cite{wwprd83014008}.
These rare B decays have been studied in the naive factorization
\cite{epjc22683,epjc22695,prd67014002,jpg36095004,arxiv1004.1928}.
Due to the fact that $\langle 0\mid j^{\mu}\mid T \rangle = 0$,
where $j^{\mu}$ is the $(V\pm A)$ or $(S\pm P)$ current
\cite{zheng1,zheng2,epjc22683,epjc22695}, the factorizable amplitude
with a tensor meson emitted vanishes. Therefore, the naive
factorization approach  for this kind of decays can not give the
right prediction. The recent developed QCD factorization approach
\cite{zheng1,zheng2} and the perturbative QCD factorization approach
(PQCD) \cite{zou} overcome these shortcoming by including the large
non-factorization contributions and the annihilation type
contributions.

There is another category of B decays with a heavy $D$ meson and a
tensor meson in the final states,  which are discussed in the
factorization approach
\cite{prd491645,prd555581,prd59077504,arxiv1010.3077,prd67014011,prd85051301}.
These $B$ decays   include the Cabibbo-Kobayashi-Maskawa-
(CKM-)favored $B$ decays through $b\rightarrow c$ transition, and
the CKM-suppressed $B$ decays through $b\rightarrow u$ transition.
There are  only tree operator contributions,  thus no CP asymmetry
appears in the standard model for these decays. Again, the
factorizable diagrams with a tensor meson emitted vanish in the
naive factorization.  To deal with the large non-factorizable
contribution and annihilation type contribution, one has to go
beyond the naive factorization.

Recently, three pure annihilation type decays $B^0\rightarrow
D_{s}^{-}K_{2}^{*+}$ and $B_{s}\rightarrow \bar{D}a_{2}$ are
calculated in the perturbative QCD approach, which give sizable
branching ratios \cite{zou-zhou}. In this work, we shall extend the
study to all the charmed $B(B_{s})\,\rightarrow\,D_{(s)}^{(*)}(\bar{D}_{(s)}^{(*)})\,T$
decays in the PQCD approach, which is based on the $k_{T}$
factorization \cite{wang7,prd63074009}. We know that the light quark
in B meson is soft, while it is collinear in the final state light
meson, so a hard gluon is necessary to connect the spectator quark
to the four quark operator. So the hard part of the PQCD approach
contains six quarks rather than four quarks. This is called
six-quark effective theory or six-quark operator. In the calculation
of the factorizable diagrams and the annihilation type diagrams,
endpoint singularity will appear to spoil the perturbative
calculation. In the conventional collinear factorization, people
usually parameterize these singularity, thus makes the theoretical
prediction weak. In the PQCD approach, the quarks' intrinsic
transverse momenta are kept to avoid the endpoint divergence.
Because of the additional energy scale introduced by the transverse
momentum, double logarithms will appear in the QCD radiative
corrections. We resum these double logarithms to give a Sudakov
factor, which effectively suppresses the end-point region
contribution. This makes the PQCD approach more reliable and
consistent. So  in the perturbative QCD approach, one can not only
give predictions for the decays with a tensor meson emitted but also
calculate the pure annihilation type B decays
\cite{annihilation1,annihilation2}.

In charmed B decays, there is one more intermediate energy scale,
the heavy D meson mass. As a result, another expansion series of
$m_{D}/m_{B}$ will appear. The factorization is only approved at the
leading of $m_{D}/m_{B}$ expansion \cite{fd1,fd2}. It is also proved
factorization in the soft collinear effective theory for this kind
of decays \cite{scet}. Therefore, we will take only the leading
order contribution into account, unless explicitly mentioned.

This paper is organized as following. In Section II, we present the
formalism and wave functions of the considered B meson decays. Then
we perform the perturbative calculations for considered decay
channels with the PQCD approach in Sec.III. The numerical results
and phenomenological analysis are given in Sec.IV. Finally, Sec.V is
a short summary.

\section{FORMALISM AND WAVE FUNCTION}\label{sec:function}

The $B\,\rightarrow\,DT$ decays are weak decays through charged
currents. At the quark level, there are only tree operator
contributions, and the related weak effective Hamiltonian $H_{eff}$
\cite{rmp68} can be written as
\begin{eqnarray}
H_{eff}=\frac{G_{F}}{\sqrt{2}}\,V_{ub}^{*}V_{cd(s)}\left[C_{1}(\mu)O_{1}(\mu)\,+\,C_{2}(\mu)O_{2}(\mu)\right],
\end{eqnarray}
where $V_{ub}$ and $V_{cd(s)}$ are CKM matrix elements. $C_{1,2}(\mu)$ are the Wilson
coefficients at the renormalization scale $\mu$. $O_{1,2}(\mu)$ are
the four quark operators.
\begin{eqnarray}
O_{1}\,=\,(\bar{b}_{\alpha}u_{\beta})_{V-A}(\bar{c}_{\beta}d(s)_{\alpha})_{V-A},
\;O_{2}\,=\,(\bar{b}_{\alpha}u_{\alpha})_{V-A}(\bar{c}_{\beta}d(s)_{\beta})_{V-A},
\end{eqnarray}
where $\alpha$ and $\beta$ are the color indices,
$(\bar{b}_{\alpha}u_{\beta})_{V-A}\,=\,\bar{b}_{\alpha}\gamma^{\mu}(1-\gamma^{5})u_{\beta}$.
Conventionally, we define the combined Wilson coefficients as
\begin{eqnarray}
a_{1}=C_{2}+C_{1}/3,\;a_{2}=C_{1}+C_{2}/3.
\end{eqnarray}
For the $B\,\rightarrow\,\bar{D}T$ decays, the decay rates will be
enhanced comparing with the corresponding $B\,\rightarrow\,DT$
decays by CKM matrix elements $|V_{cb}/V_{ub}|^2$. At quark level,
these decays are governed by the effective Hamiltonian
\begin{eqnarray}
H_{eff}=\frac{G_{F}}{\sqrt{2}}\,V_{cb}^{*}V_{ud(s)}\left[C_{1}(\mu)O_{1}(\mu)\,+\,C_{2}(\mu)O_{2}(\mu)\right],
\end{eqnarray}
with
\begin{eqnarray}
O_{1}\,=\,(\bar{b}_{\alpha}c_{\beta})_{V-A}(\bar{u}_{\beta}d(s)_{\alpha})_{V-A},
\;O_{2}\,=\,(\bar{b}_{\alpha}c_{\alpha})_{V-A}(\bar{u}_{\beta}d(s)_{\beta})_{V-A}.
\end{eqnarray}

In hadronic $B$ decay calculations, one has to deal with the
hadronization of mesons. In this calculation, there are three
different scales: W boson mass scale, b quark mass scale $M_{B}$ and
the factorization scale $\sqrt{\bar{\Lambda}M_{B}}$, where
$\bar{\Lambda}\equiv M_{B}-m_{b}$. The electroweak physics higher
than W boson mass can be calculated perturbatively. The physics
between b quark mass scale and W boson mass scale can be included in
the Wilson coefficients of the effective
 four-quark operators, which is obtained by using the renormalization group equation. The physics between $M_{B}$ and the
  factorization scale is included in the calculation of the
   hard part in the PQCD approach. The physics below the factorization scale
   is nonperturbative and described by the hadronic wave functions of mesons, which is universal for all decay modes.
Therefore,  the decay amplitude can be explicitly factorized into
the convolution of the the Wilson coefficients, the hard scattering
kernel and the light-cone wave functions of mesons characterized by
different scales, respectively,
\begin{eqnarray}
\mathcal
{A}\;\sim\;&&\int\,dx_{1}dx_{2}dx_{3}b_{1}db_{1}b_{2}db_{2}b_{3}db_{3}\nonumber\\
&&\times
Tr\left[C(t)\Phi_{B}(x_{1},b_{1})\Phi_{M_{2}}(x_{2},b_{2})\Phi_{M_{3}}(x_{3},b_{3})H(x_{i},b_{i},t)S_{t}(x_{i})e^{-S(t)}\right],
\end{eqnarray}
where  $b_{i}$ is the conjugate variable of quark's transverse
momentum $k_{iT}$, $x_{i}$ is the momentum fractions of valence
quarks, and $t$ is the largest energy scale in the hard part
$H(x_{i},b_{i},t)$. $C(t)$ are the Wilson coefficients with
resummation of the large logarithms $\ln(m_{W}/t)$ produced by the
radiative corrections. $S_{t}(x_{i})$ is the jet function, which is
obtained by the threshold resummation and smears the end-point
singularities on $x_{i}$ \cite{prd66094010}. The last term,
$e^{-S(t)}$, is the Sudakov form factor which suppresses the soft
dynamics effectively and suppresses the long distance contributions
in the large $b$ region \cite{prd57443,lvepjc23275}. Thus it makes
the perturbative calculation of the hard part $H$ applicable at
intermediate scale, i.e., $m_{B}$ scale.


In the PQCD approach,  in order to calculate the decay amplitude, we
should choose the proper wave functions for the initial and final
state mesons. The initial B meson is a heavy pseudoscalar  meson
with two Lorentz structures in it's wave function. We have neglected
the numerically-suppressed one in the PQCD approach
\cite{epjc28515}. The two rest structure ($\gamma_{\mu}\gamma_{5}$)
and $\gamma_{5}$ components remain as leading contributions
\cite{wwprd83014008}. Then, $\Phi_{B}$ can be written as
\begin{eqnarray}
\Phi_{B}=\frac{i}{\sqrt{6}}\left[\left(\makebox[-1.5pt][l]{/}P+m_{B}\right)\gamma_{5}\phi_{B}(x)\right].
\end{eqnarray}
For the distribution amplitude, we  choose
\cite{prd65014007,epjc28515}
\begin{eqnarray}
\phi_{B}(x,b)=N_{B}x^{2}(1-x)^{2}\exp\left[-\frac{1}{2}\left(\frac{m_{B}x}{\omega_{B}}\right)^{2}\,-\,\frac{\omega_{B}^{2}b^{2}}{2}\right],
\end{eqnarray}
where $N_{B}$ is the normalization constant. We will take
$\omega_{B}=(0.4\,\pm\,0.04)$ GeV and $f_{B}=(0.21\,\pm\,0.02)$ GeV
for B meson \cite{zheng1,zheng2,wang7,prd65014007,wang13,epjc28515}.
For the $B_{s}$ meson, because of the SU(3) breaking effects, we
choose $\omega_{B}\,=\,(0.5\pm0.05)$ GeV \cite{wang14} and
$f_{B_{s}}\,=\,(0.24\pm0.03)$ GeV.

For a tensor meson, the polarization tensor
$\epsilon_{\mu\nu}(\lambda)$ with helicity $\lambda$ can be expanded
through the polarization vectors $\epsilon^{\mu}(0)$ and
$\epsilon^{\mu}(\pm1)$ \cite{zheng1,zheng2}
\begin{eqnarray}
\epsilon^{\mu\nu}(\pm2)\,&\equiv&\,\epsilon(\pm1)^{\mu}\epsilon(\pm1)^{\nu},\nonumber\\
\epsilon^{\mu\nu}(\pm1)\,&\equiv&\,\sqrt{\frac{1}{2}}\left[\epsilon(\pm1)^{\mu}\epsilon(0)^{\nu}\,
+\,\epsilon(0)^{\mu}\epsilon(\pm1)^{\nu}\right],\nonumber\\
\epsilon^{\mu\nu}(0)\,&\equiv&\,\sqrt{\frac{1}{6}}\left[\epsilon(+1)^{\mu}\epsilon(-1)^{\nu}\,
+\,\epsilon(-1)^{\mu}\epsilon(+1)^{\nu}\right]\,+\,\sqrt{\frac{2}{3}}\epsilon(0)^{\mu}\epsilon(0)^{\nu}.
\end{eqnarray}
 In order to calculate
conveniently, we define a new polarization vector $\epsilon_{T}$ for
the considered tensor meson \cite{wwprd83014008}
\begin{eqnarray}
\epsilon_{T\mu}=\frac{1}{m_{B}}\epsilon_{\mu\nu}(h)P_{B}^{\nu}.
\end{eqnarray}
The new polarization vector $\epsilon_{T}(\lambda)$ with helicity
$\lambda$ can be expressed as
\begin{eqnarray}
&&\epsilon_{T\mu}(\pm2)=0,\nonumber\\
&&\epsilon_{T\mu}(\pm1)=\frac{1}{m_{B}}\frac{1}{\sqrt{2}}\,\left(\epsilon(0)\cdot
P_{B}\right)\,\epsilon_{\mu}(\pm1),\nonumber\\
&&\epsilon_{T\mu}(0)=\frac{1}{m_{B}}\sqrt{\frac{2}{3}}\,\left(\epsilon(0)\cdot
P_{B}\right)\,\epsilon_{\mu}(0). \label{vector}
\end{eqnarray}
 In this
convention, the $\pm 2$ polarizations do not contribute, which is
consistent with the angular momentum conservation argument in B
decays. The $\epsilon_{T}$ is similar with the $\epsilon$ of vector state,
regardless of the related constants \cite{wwprd83014008}. This convention makes the following perturbative
calculations simpler. After this simplification,
 the wave function for a generic
tensor meson are defined by \cite{wwprd83014008}
\begin{eqnarray}
&&\Phi_{T}^{L}\,=\,\frac{1}{\sqrt{6}}\left[m_{T}\makebox[0pt][l]{/}\epsilon_{\bullet
L}^{*}\phi_{T}(x)\,+\,\makebox[0pt][l]{/}\epsilon_{\bullet
L}^{*}\makebox[-1.5pt][l]{/}P\phi_{T}^{t}(x)+m_{T}^{2}\frac{\epsilon_{\bullet}\cdot
v}{P\cdot v}\phi_{T}^{s}(x)\right]\nonumber\\
&&\Phi_{T}^{\perp}\,=\,\frac{1}{\sqrt{6}}\left[m_{T}\makebox[0pt][l]{/}\epsilon_{\bullet
\perp}^{*}\phi_{T}^{v}(x)\,+\,\makebox[0pt][l]{/}\epsilon_{\bullet
\perp}^{*}\makebox[-1.5pt][l]{/}P\phi_{T}^{T}(x)\,+\,m_{T}i\epsilon_{\mu\nu\rho\sigma}\gamma_{5}\gamma^{\mu}\epsilon_{\bullet
\perp}^{* \nu}n^{\rho}v^{\sigma}\phi_{T}^{a}(x)\right].
\end{eqnarray}
Here $n$ is the moving direction of the tensor meson, and $v$ is the
opposite direction. We adopt the convention $\epsilon^{0123}=1$. The
vector
$\epsilon_{\bullet\mu}\,\equiv\,\frac{\epsilon_{\mu\nu}v^{\nu}}{P\cdot\,
v}$ is related to the polarization tensor. The twist-2 and twist-3
distribution amplitudes are given by
\cite{wwprd83014008,zheng1,zheng2}
\begin{eqnarray}
&&\phi_{T}(x)\,=\,\frac{f_{T}}{2\sqrt{2N_{c}}}\phi_{\|}(x),\;\phi_{T}^{t}\,=\,\frac{f_{T}^{\perp}}{2\sqrt{2N_{c}}}h_{\|}^{(t)}(x),
\nonumber\\
&&\phi_{T}^{s}(x)\,=\,\frac{f_{T}^{\perp}}{4\sqrt{2N_{c}}}\frac{d}{dx}h_{\|}^{(s)}(x),\;
\phi_{T}^{T}(x)\,=\,\frac{f_{T}^{\perp}}{2\sqrt{2N_{c}}}\phi_{\perp}(x),\nonumber\\
&&\phi_{T}^{v}(x)\,=\,\frac{f_{T}}{2\sqrt{2N_{C}}}g_{\perp}^{(v)}(x),\;\phi_{T}^{a}(x)\,=
\,\frac{f_{T}}{8\sqrt{2N_{c}}}\frac{d}{dx}g_{\perp}^{(a)}(x),
\end{eqnarray}
with the form
\begin{eqnarray}
&&\phi_{\|,\perp}(x)\,=\,30x(1-x)(2x-1),\nonumber\\
&&h_{\|}^{(t)}(x)\,=\,\frac{15}{2}(2x-1)(1-6x+6x^{2}),\;h_{\|}^{(s)}(x)\,=\,15x(1-x)(2x-1),\nonumber\\
&&g_{\perp}^{(a)}(x)\,=\,20x(1-x)(2x-1),\;\;g_{\perp}^{(v)}(x)\,=\,5(2x-1)^{3}.
\end{eqnarray}
It is obvious that all  the above light-cone distribution
 amplitudes (LCDAs) of the tensor meson are antisymmetric under the
 interchange of momentum fractions of the quark and anti-quark in
 the SU(3) limit (i.e. $x\leftrightarrow 1-x$) \cite{zheng1,zheng2}. This is required by  the Bose statistics,
 and consistent with the
 fact that  $<0\mid j^{\mu}\mid T>\,=\,0$, where $j^{\mu}$
is the $(V\pm A)$ or $(S\pm P)$ current.

 For $D^{(*)}$ meson, in the
heavy quark limit, the two-parton LCDAs can be written as
\cite{prd67054028,liprd78014018,zouhaojpg37,liprd81034006}
\begin{eqnarray}
\langle D(p)|q_{\alpha}(z)\bar{c}_{\beta}(0)|0\rangle
\,&=&\,\frac{i}{\sqrt{2N_{c}}}\int_{0}^{1}dx\,e^{ixp\cdot
z}\left[\gamma_{5}(\makebox[-1.5pt][l]{/}P\,+\,m_{D})\phi_{D}(x,b)\right]_{\alpha\beta},\nonumber\\
\langle D^{*}(p)|q_{\alpha}(z)\bar{c}_{\beta}(0)|0\rangle
\,&=&\,-\frac{1}{\sqrt{2N_{c}}}\int_{0}^{1}dx\,e^{ixp\cdot
z}\left[\makebox[-1.5pt][l]{/}\epsilon_{L}(\makebox[-1.5pt][l]{/}P\,+\,m_{D^{*}})\phi_{D^{*}}^{L}(x,b)\right.\nonumber\\
&&\left.\,+\;\makebox[-1.5pt][l]{/}\epsilon_{T}(\makebox[-1.5pt][l]{/}P\,+\,m_{D^{*}})\phi_{D^{*}}^{T}(x,b)\right]_{\alpha\beta}.
\end{eqnarray}
 For the distribution amplitude for $D^{(*)}$ meson, we take the same
 as that used in Refs.
\cite{liprd78014018,zouhaojpg37,liprd81034006}.
\begin{eqnarray}
\phi_{D}(x,b)\,=\,\phi_{D^{*}}^{L(T)}(x,b)=\frac{1}{2\sqrt{2N_{c}}}\,f_{D^{(*)}}\,6x(1-x)\left[1+C_{D}(1-2x)\right]
\exp\left[\frac{-\omega^{2}b^{2}}{2}\right],
\end{eqnarray}
with $C_{D}=0.5\pm0.1, \omega=0.1$ GeV and $f_{D}=207$ MeV \cite{fd}
for $D (\bar{D}$) meson and $C_{D}=0.4\pm0.1, \omega=0.2$ GeV and
$f_{D_{s}}=241$ MeV \cite{fd}  for $D_{s} (\bar{D}_{s})$ meson. For
$D_{(s)}^{*}$ meson, we determine the decay constant of $D_{(s)}^{*}$ meson by
using the following relation based on heavy quark effective theory
(HQET) \cite{am}
\begin{eqnarray}
f_{D_{(s)}^{*}}\,=\,\sqrt{\frac{m_{D_{(s)}}}{m_{D_{(s)}}^{*}}}\,f_{D_{(s)}}.
\end{eqnarray}

\section{Perturbative calculation}\label{sec:bcdv}

In this section, we shall calculate the hard part $H(t)$, which is
decay channel dependent.  It includes the four quark operators and
the necessary hard gluon connecting the four quark operator and the
spectator quark \cite{lvepjc23275}. We will express the whole
amplitude for each diagram as the convolution of the hard kernal and
wave functions.

\begin{figure}[]
\begin{center}
\vspace{-5cm} \centerline{\epsfxsize=10 cm \epsffile{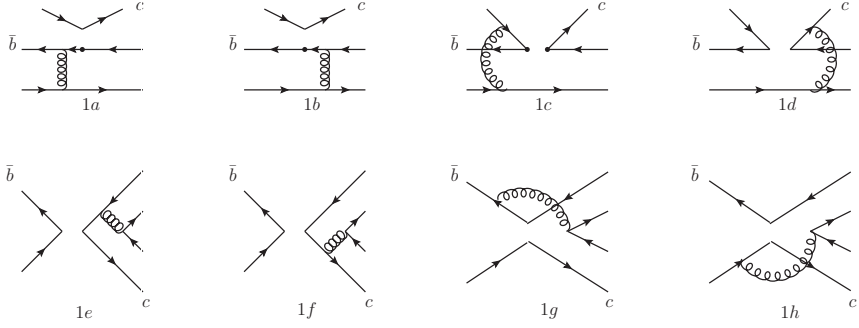}}
\vspace{-4cm} \caption{Leading order Feynman diagrams contributing to the
$B\,\rightarrow\,D^{(*)}T$ decays in PQCD}
 \label{fig:lodiagram}
 \end{center}
\end{figure}

There are 8 types of diagrams contributing to the  $ B \rightarrow D
T$ decays ($\bar{b}\rightarrow c$ transition), which are shown in
Fig.1. They are governed by the CKM matrix element $V_{ub}$, which
are usually called CKM suppressed decay channels. The first two
diagrams of Fig.1 are the factorizable diagrams. Their decay
amplitude can be factorized as a product of the decay constant of
$D^{(*)}$ meson and a B to tensor meson transition form factor in
the naive factorization approach. In the PQCD approach, we calculate
these two
 diagrams and obtain the decay amplitude
 as
\begin{eqnarray}
\mathcal {A}_{ef}\,=&&\,-8\sqrt{\frac{2}{3}}\pi
C_{F}m_{B}^{4}f_{D}\int_{0}^{1}dx_{1}dx_{3}\int_{0}^{1/\Lambda}b_{1}db_{1}b_{3}db_{3}\phi_{B}(x_{1},b_{1})\nonumber\\
&&\times\left\{\left[\phi_{T}(x_{3})(x_{3}+1)-(\phi_{T}^{s}(x_{3})+\phi_{T}^{t}(x_{3}))r_{T}(2x_{3}-1)\right]\right.\nonumber\\
&&\left.\cdot
E_{ef}(t_{a})h_{ef}(x_{1},x_{3}(1-r_{D}^{2}),b_{1},b_{3})\right.\nonumber\\
&&\left.+2r_{T}\phi_{T}^{s}(x_{3})E_{ef}(t_{b})h_{ef}(x_{3},x_{1}(1-r_{D}^{2}),b_{3},b_{1})\right\},
\label{ef}
\end{eqnarray}
with  $r_{T}=\frac{m_{T}}{m_{B}}$ and $r_{D}=\frac{m_{D}}{m_{B}}$.
$\phi_{T}^{(s,t)}(x_{i})$ is the distribution amplitude of the
tensor meson and $C_{F}\,=\,\frac{4}{3}$ is a color factor.  The functions $h_{ef}$,
$t_{a,b}$, $S_{t}$ and $E_{ef}$ can be found in Appendix \ref{a}.

For diagrams Fig.(1c) and (1d), which are the non-factorizable in naive factorization, the decay
amplitudes involve all three meson wave functions. The integration
of $b_{3}$ can be performed through $\delta$ function
$\delta(b_{1}-b_{3})$, leaving only integration of $b_{1}$ and
$b_{2}$.
\begin{eqnarray}
\mathcal {M}_{enf}&=&-\frac{32}{3}C_{F}\pi
m_{B}^{4}\int_{0}^{1}\,dx_{1}dx_{2}dx_{3}\int_{0}^{1/\Lambda}\,b_{1}db_{1}b_{2}db_{2}\,\phi_{B}(x_{1},b_{1})\phi_{D}(x_{2},b_{2})\nonumber\\
&&\times
\left\{\left[\phi_{T}(x_{3})x_{2}+(\phi_{T}^{t}(x_{3})-\phi_{T}^{s}(x_{3}))r_{T}x_{3}\right]\right.\nonumber\\
&&\cdot \left. h_{enf1}(x_{i},b_{i})E_{enf}(t_{c})\right.\nonumber\\
&&\left.+\left[\phi_{T}(x_{3})(x_{2}-x_{3}-1)+(\phi_{T}^{s}(x_{3})+\phi_{T}^{t}(x_{3}))r_{T}x_{3}\right]\right.\nonumber\\
&& \cdot \left. h_{enf2}(x_{i},b_{i})E_{enf}(t_{d})\right\}.
\label{enf}
\end{eqnarray}

For the factorizable annihilation type diagrams Fig.(1e) and (1f),
the decay amplitudes involve only the final state meson wave
functions, with the B meson
 factorized out,
\begin{eqnarray}
\mathcal {A}_{af}&=&8\sqrt{\frac{2}{3}}C_{F}f_{B}\pi
m_{B}^{4}\int_{0}^{1}\,dx_{2}dx_{3}\int_{0}^{1/\Lambda}\,b_{2}db_{2}b_{3}db_{3}\,\phi_{D}(x_{3},b_{3})\nonumber\\
&&\times\left\{\left[\phi_{T}(x_{2})x_{3}+2r_{D}r_{T}\phi_{T}^{s}(x_{2})(x_{3}+1)\right]\right.\nonumber\\
&&\left.\cdot h_{af}(x_{2},x_{3}(1-r_{D}^{2}),b_{2},b_{3})E_{af}(t_{e})\right.\nonumber\\
&&\left.-\left[\phi_{T}(x_{2})x_{2}+r_{D}r_{T}(\phi_{T}^{t}(x_{2})(2x_{2}-1)+\phi_{T}^{s}(x_{2})(2x_{2}+1))\right]\right.\nonumber\\
&&\cdot\left.h_{af}(x_{3},x_{2}(1-r_{D}^{2}),b_{3},b_{2})E_{af}(t_{f})\right\}.
\label{af}
\end{eqnarray}
For the non-factorizable annihilation diagrams Fig.(1g) and (1h),
all three meson wave functions are involved in the decay amplitudes.
The integration of $b_{3}$ can be performed by the $\delta$ function
$\delta(b_{2}-b_{3})$ to give the decay amplitudes as
\begin{eqnarray}
\mathcal {M}_{anf}&=&\frac{32}{3}C_{F}\pi
m_{B}^{4}\int_{0}^{1}\,dx_{1}dx_{2}dx_{3}\int_{0}^{1/\Lambda}\,b_{1}db_{1}b_{2}db_{2}\,\phi_{B}(x_{1},b_{1})\phi_{D}(x_{3},b_{2})\nonumber\\
&&\times \left\{\left[\phi_{T}(x_{2})x_{2}+r_{D}r_{T}(\phi_{T}^{t}(x_{2})(x_{2}-x_{3})+\phi_{T}^{s}(x_{2})(x_{2}+x_{3}+2))\right]\right.\nonumber\\
&&\left.\cdot h_{anf1}(x_{1},x_{2},x_{3},b_{1},b_{2})E_{anf}(t_{g})\right.\nonumber\\
&&\left.-\left[\phi_{T}(x_{2})x_{3}+r_{D}r_{T}(\phi_{T}^{t}(x_{2})(x_{3}-x_{2})+\phi_{T}^{s}(x_{2})(x_{2}+x_{3}))\right]\right.\nonumber\\
&&\cdot\left.h_{anf2}(x_{1},x_{2},x_{3},b_{1},b_{2})E_{anf}(t_{h})\right\}.
\end{eqnarray}

 The situation for $B\rightarrow D^{*}(\bar{D}^{*}) T$ mode is a little more
complicated. Both the longitudinal polarization and the transverse
polarization contribute.   Their decay amplitude can be given by
\begin{eqnarray}
\mathcal {A}(\epsilon_{D},\epsilon_{T})=i\mathcal
{A}^{L}+i(\epsilon_{D}^{T*}\cdot\epsilon_{T}^{T*})\mathcal
{A}^{T}+(\epsilon_{\mu\nu\alpha\beta}n^{\mu}v^{\nu}\epsilon_{D}^{T*\alpha}\epsilon_{T}^{T*\beta})\mathcal
{A}^{N},
\end{eqnarray}
where $ \mathcal {A}^{L}$ is the longitudinally polarized decay
amplitude and $ \mathcal {A}^{T}$ and $ \mathcal {A}^{N}$ are the
transversely polarized contributions. $\epsilon_{D}^{T}$ is
 the transverse polarization vector
of $D^{*}(\bar{D}^{*})$ and $\epsilon_{T}^{T}$ is the vector used to construct the
polarization tensors of the tensor meson.

For $B\rightarrow D^{*}T$ decay mode, the longitudinally polarized
expressions of factorizable and nonfactorizable emission
contributions can be obtained by making the following substitutions
in Eq.(\ref{ef}) and Eq.(\ref{enf}),
\begin{eqnarray}
\phi_{D}\rightarrow \phi_{D^{*}}^{L},\;f_{D}\rightarrow
f_{D^{*}},\;m_{D}\rightarrow m_{D^{*}}. \label{tihuan}
\end{eqnarray}
For annihilation type diagrams, the longitudinal decay amplitudes
are
\begin{eqnarray}
\mathcal {A}_{af}^{L}&=&8\sqrt{\frac{2}{3}}C_{F}f_{B}\pi
m_{B}^{4}\int_{0}^{1}\,dx_{2}dx_{3}\int_{0}^{1/\Lambda}\,b_{2}db_{2}b_{3}db_{3}\,\phi_{D}^{L}(x_{3},b_{3})\nonumber\\
&&\times\left\{\left[\phi_{T}(x_{2})x_{3}+2r_{D}r_{T}\phi_{T}^{s}(x_{2})(x_{3}-1)\right]\right.\nonumber\\
&&\left.\cdot h_{af}(x_{2},x_{3}(1-r_{D}^{2}),b_{2},b_{3})E_{af}(t_{e})\right.\nonumber\\
&&\left.-\left[\phi_{T}(x_{2})x_{2}+r_{D}r_{T}(\phi_{T}^{s}(x_{2})-\phi_{T}^{t}(x_{2}))\right]\right.\nonumber\\
&&\cdot\left.h_{af}(x_{3},x_{2}(1-r_{D}^{2}),b_{3},b_{2})E_{af}(t_{f})\right\},
\end{eqnarray}
\begin{eqnarray}
\mathcal {M}_{anf}^{L}&=&\frac{32}{3}C_{F}\pi
m_{B}^{4}\int_{0}^{1}\,dx_{1}dx_{2}dx_{3}\int_{0}^{1/\Lambda}\,b_{1}db_{1}b_{2}db_{2}\,\phi_{B}(x_{1},b_{1})\phi_{D}^{L}(x_{3},b_{2})\nonumber\\
&&\times \left\{\left[\phi_{T}(x_{2})x_{2}+r_{D}r_{T}(\phi_{T}^{s}(x_{2})(x_{2}-x_{3})+\phi_{T}^{t}(x_{2})(x_{2}+x_{3}-2))\right]\right.\nonumber\\
&&\left.\cdot h_{anf1}(x_{1},x_{2},x_{3},b_{1},b_{2})E_{anf}(t_{g})\right.\nonumber\\
&&\left.+\left[-\phi_{T}(x_{2})x_{3}+r_{D}r_{T}(\phi_{T}^{s}(x_{2})(x_{2}-x_{3})-\phi_{T}^{t}(x_{2})(x_{2}+x_{3}))\right]\right.\nonumber\\
&&\cdot\left.h_{anf2}(x_{1},x_{2},x_{3},b_{1},b_{2})E_{anf}(t_{h})\right\}.
\end{eqnarray}
The transversely polarized contributions are suppressed by $r_{D}$
or $r_{T}$, whose decay amplitudes can be given by
\begin{eqnarray}
\mathcal {A}_{ef}^{T}\,=&&\,-4\sqrt{2}\pi
C_{F}m_{B}^{4}f_{D^{*}}r_{D}\int_{0}^{1}dx_{1}dx_{3}\int_{0}^{1/\Lambda}b_{1}db_{1}b_{3}db_{3}\phi_{B}(x_{1},b_{1})\nonumber\\
&&\times\left\{\left[\phi_{T}^{T}(x_{3})+r_{T}(\phi_{T}^{v}(x_{3})(x_{3}+2)-\phi_{T}^{a}(x_{3})x_{3})\right]\right.\nonumber\\
&&\left.\cdot
E_{ef}(t_{a})h_{ef}(x_{1},x_{3}(1-r_{D}^{2}),b_{1},b_{3})\right.\nonumber\\
&&\left.+r_{T}(\phi_{T}^{a}(x_{3})+\phi_{T}^{v}(x_{3}))E_{ef}(t_{b})h_{ef}(x_{3},x_{1}(1-r_{D}^{2}),b_{3},b_{1})\right\},
\label{eft}
\end{eqnarray}
\begin{eqnarray}
\mathcal {A}_{ef}^{N}\,=\,\mathcal
{A}_{ef}^{T}(\phi_{T}^{a}\leftrightarrow \phi_{T}^{v}),
\label{efn}
\end{eqnarray}
\begin{eqnarray}
\mathcal {M}_{enf}^{T}&=&16\sqrt{\frac{1}{3}}C_{F}\pi
m_{B}^{4}r_{D}\int_{0}^{1}\,dx_{1}dx_{2}dx_{3}\int_{0}^{1/\Lambda}\,b_{1}db_{1}b_{2}db_{2}\,\phi_{B}(x_{1},b_{1})\phi_{D}^{T}(x_{2},b_{2})\nonumber\\
&&\times
\left\{\left[-\phi_{T}^{T}(x_{3})x_{2}\right]h_{enf1}(x_{i},b_{i})E_{enf}(t_{c})\right.\nonumber\\
&&\left.+\left[\phi_{T}^{T}(x_{3})(x_{2}+1)+r_{T}(\phi_{T}^{v}(x_{3})(1-2x_{2}+2x_{3})-\phi_{T}^{a}(x_{3}))\right]\right.\nonumber\\
&& \cdot \left. h_{enf2}(x_{i},b_{i})E_{enf}(t_{d})\right\},
\end{eqnarray}
\begin{eqnarray}
\mathcal {M}_{enf}^{N}\,=\,\mathcal
{M}_{enf}^{T}(\phi_{T}^{a}\leftrightarrow \phi_{T}^{v}),
\end{eqnarray}
\begin{eqnarray}
\mathcal {A}_{af}^{T}&=&4\sqrt{2}C_{F}f_{B}\pi
m_{B}^{4}r_{D}\int_{0}^{1}\,dx_{2}dx_{3}\int_{0}^{1/\Lambda}\,b_{2}db_{2}b_{3}db_{3}\,\phi_{D}^{T}(x_{3},b_{3})\nonumber\\
&&\times\left\{r_{T}\left[\phi_{T}^{a}(x_{2})(1-x_{3})+\phi_{T}^{v}(x_{2})(x_{3}+1)\right]\right.\nonumber\\
&&\left.\cdot h_{af}(x_{2},x_{3}(1-r_{D}^{2}),b_{2},b_{3})E_{af}(t_{e})\right.\nonumber\\
&&\left.+\left[r_{D}\phi_{T}^{T}(x_{2})+r_{T}(\phi_{T}^{a}(x_{2})(1-x_{2})-\phi_{T}^{v}(x_{2})(x_{2}+1))\right]\right.\nonumber\\
&&\cdot\left.h_{af}(x_{3},x_{2}(1-r_{D}^{2}),b_{3},b_{2})E_{af}(t_{f})\right\},
\end{eqnarray}
\begin{eqnarray}
\mathcal {A}_{af}^{N}\,=\,-\mathcal
{A}_{af}^{T}(\phi_{T}^{a}\leftrightarrow \phi_{T}^{v}),
\end{eqnarray}
\begin{eqnarray}
\mathcal {M}_{anf}^{T}&=&16\sqrt{\frac{1}{3}}\,C_{F}\pi
m_{B}^{4}r_{D}\int_{0}^{1}\,dx_{1}dx_{2}dx_{3}\int_{0}^{1/\Lambda}\,b_{1}db_{1}b_{2}db_{2}\,\phi_{B}(x_{1},b_{1})\phi_{D}^{T}(x_{3},b_{2})\nonumber\\
&&\times \left\{\left[2r_{T}\phi_{T}^{v}(x_{2})+r_{D}\phi_{T}^{T}(x_{2})(x_{3}-1)\right]h_{anf1}(x_{1},x_{2},x_{3},b_{1},b_{2})E_{anf}(t_{g})\right.\nonumber\\
&&\left.-\left[r_{D}x_{3}\phi_{T}^{T}(x_{2})\right]h_{anf2}(x_{1},x_{2},x_{3},b_{1},b_{2})E_{anf}(t_{h})\right\}.
\end{eqnarray}
\begin{eqnarray}
\mathcal {M}_{anf}^{N}\,=\,-\mathcal
{M}_{anf}^{T}(\phi_{T}^{v}\rightarrow \phi_{T}^{a}).
\end{eqnarray}

\begin{figure}[]
\begin{center}
\vspace{-5cm} \centerline{\epsfxsize=10 cm \epsffile{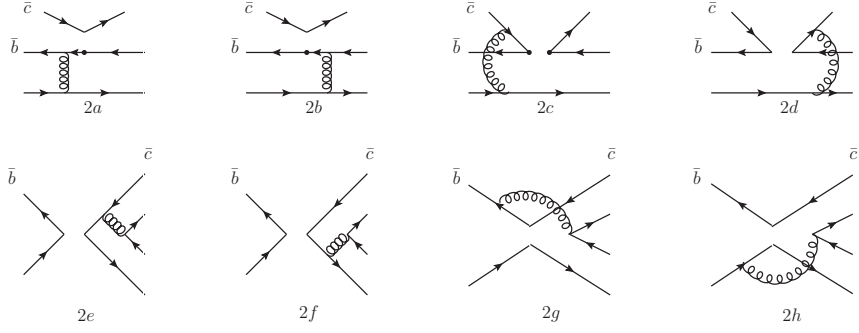}}
\vspace{-4cm} \caption{Leading order Feynman diagrams contributing to the
$B\,\rightarrow\,\bar{D}T$ decays}
 \label{fig:lodiagram2}
 \end{center}
\end{figure}

The complete decay amplitudes of each $B_{(s)}\rightarrow D_{(s)}T$
channels are then
\begin{eqnarray}
\mathcal {A}(B^{0}\rightarrow
D^{0}a_{2}^{0})&=&\frac{G_{F}}{\sqrt{2}}\frac{1}{\sqrt{2}}V_{ub}^{*}V_{cd}\left[a_{2}\mathcal
{A}_{af}+C_{2}\mathcal {M}_{anf}-a_{2}\mathcal
{A}_{ef}-C_{2}\mathcal {M}_{enf}\right],\\
 \mathcal
{A}(B^{0}\rightarrow
D^{0}f_{2}^{q})\,&=&\,\frac{G_{F}}{\sqrt{2}}\frac{1}{\sqrt{2}}V_{ub}^{*}V_{cd}\left[a_{2}\mathcal
{A}_{ef}+C_{2}\mathcal {M}_{enf}+a_{2}\mathcal
{A}_{af}+C_{2}\mathcal {M}_{anf}\right],\\
 \mathcal
{A}(B^{0}\rightarrow
D^{0}K_{2}^{*0})\,&=&\,\frac{G_{F}}{\sqrt{2}}V_{ub}^{*}V_{cs}\left[a_{2}\mathcal
{A}_{ef}+C_{2}\mathcal {M}_{enf}\right],\\
\mathcal {A}(B^{0}\rightarrow
D^{+}a_{2}^{-})&=&\frac{G_{F}}{\sqrt{2}}V_{ub}^{*}V_{cd}\left[a_{1}\mathcal
{A}_{ef}+C_{1}\mathcal {M}_{enf}+a_{2}\mathcal
{A}_{af}+C_{2}\mathcal {M}_{anf}\right],\\
\mathcal {A}(B^{0}\rightarrow
D_{s}^{+}a_{2}^{-})&=&\frac{G_{F}}{\sqrt{2}}V_{ub}^{*}V_{cs}\left[a_{1}\mathcal
{A}_{ef}+C_{1}\mathcal {M}_{enf}\right],\\
\mathcal {A}(B^{0}\rightarrow
D_{s}^{+}K_{2}^{*-})&=&\frac{G_{F}}{\sqrt{2}}V_{ub}^{*}V_{cd}\left[a_{2}\mathcal
{A}_{af}+C_{2}\mathcal {M}_{anf}\right],\\
\mathcal {A}(B^{+}\rightarrow
D^{0}a_{2}^{+})&=&\frac{G_{F}}{\sqrt{2}}V_{ub}^{*}V_{cd}\left[a_{2}\mathcal
{A}_{ef}+C_{2}\mathcal {M}_{enf}+a_{1}\mathcal
{A}_{af}+C_{1}\mathcal {M}_{anf}\right],\\
\mathcal {A}(B^{+}\rightarrow
D^{0}K_{2}^{*+})&=&\frac{G_{F}}{\sqrt{2}}V_{ub}^{*}V_{cs}\left[a_{2}\mathcal
{A}_{ef}+C_{2}\mathcal {M}_{enf}+a_{1}\mathcal
{A}_{af}+C_{1}\mathcal {M}_{anf}\right]\\
\mathcal {A}(B^{+}\rightarrow
D^{+}a_{2}^{0})&=&\frac{G_{F}}{\sqrt{2}}\frac{1}{\sqrt{2}}V_{ub}^{*}V_{cd}\left[a_{1}\mathcal
{A}_{ef}+C_{1}\mathcal {M}_{enf}-a_{1}\mathcal
{A}_{af}-C_{1}\mathcal {M}_{anf}\right],\\
\mathcal {A}(B^{+}\rightarrow
D^{+}f_{2}^{q})&=&\frac{G_{F}}{\sqrt{2}}\frac{1}{\sqrt{2}}V_{ub}^{*}V_{cd}\left[a_{1}\mathcal
{A}_{ef}+C_{1}\mathcal {M}_{enf}+a_{1}\mathcal
{A}_{af}+C_{1}\mathcal {M}_{anf}\right],\\
\mathcal {A}(B^{+}\rightarrow
D^{+}K_{2}^{*0})&=&\frac{G_{F}}{\sqrt{2}}V_{ub}^{*}V_{cs}\left[a_{1}\mathcal
{A}_{af}+C_{1}\mathcal {M}_{anf}\right],\\
\mathcal {A}(B^{+}\rightarrow
D_{s}^{+}a_{2}^{0})&=&\frac{G_{F}}{\sqrt{2}}\frac{1}{\sqrt{2}}V_{ub}^{*}V_{cs}\left[a_{1}\mathcal
{A}_{ef}+C_{1}\mathcal {M}_{enf}\right],\\
\mathcal {A}(B^{+}\rightarrow
D_{s}^{+}f_{2}^{q})&=&\frac{G_{F}}{\sqrt{2}}\frac{1}{\sqrt{2}}V_{ub}^{*}V_{cs}\left[a_{1}\mathcal
{A}_{ef}+C_{1}\mathcal {M}_{enf}\right],\\
\mathcal {A}(B^{+}\rightarrow
D_{s}^{+}f_{2}^{s})&=&\frac{G_{F}}{\sqrt{2}}V_{ub}^{*}V_{cs}\left[a_{1}\mathcal
{A}_{af}+C_{1}\mathcal {M}_{anf}\right],\\
\mathcal {A}(B^{+}\rightarrow
D_{s}^{+}\bar{K}_{2}^{*0})&=&\frac{G_{F}}{\sqrt{2}}V_{ub}^{*}V_{cd}\left[a_{1}\mathcal
{A}_{af}+C_{1}\mathcal {M}_{anf}\right],\\
\mathcal {A}(B_{s}^{0}\rightarrow
D^{0}a_{2}^{0})&=&\frac{G_{F}}{\sqrt{2}}\frac{1}{\sqrt{2}}V_{ub}^{*}V_{cs}\left[a_{2}\mathcal
{A}_{af}+C_{2}\mathcal {M}_{anf}\right],\\
\mathcal {A}(B_{s}^{0}\rightarrow
D^{0}f_{2}^{q})&=&\frac{G_{F}}{\sqrt{2}}\frac{1}{\sqrt{2}}V_{ub}^{*}V_{cs}\left[a_{2}\mathcal
{A}_{af}+C_{2}\mathcal {M}_{anf}\right],\\
\mathcal {A}(B_{s}^{0}\rightarrow
D^{0}f_{2}^{s})&=&\frac{G_{F}}{\sqrt{2}}V_{ub}^{*}V_{cs}\left[a_{2}\mathcal
{A}_{ef}+C_{2}\mathcal {M}_{enf}\right],\\
\mathcal {A}(B_{s}^{0}\rightarrow
D^{0}\bar{K}_{2}^{*0})&=&\frac{G_{F}}{\sqrt{2}}V_{ub}^{*}V_{cd}\left[a_{2}\mathcal
{A}_{ef}+C_{2}\mathcal {M}_{enf}\right],
\end{eqnarray}
\begin{eqnarray}
\mathcal {A}(B_{s}^{0}\rightarrow
D^{+}a_{2}^{-})&=&\frac{G_{F}}{\sqrt{2}}V_{ub}^{*}V_{cs}\left[a_{2}\mathcal
{A}_{af}+C_{2}\mathcal {M}_{anf}\right],\\
\mathcal {A}(B_{s}^{0}\rightarrow
D^{+}K_{2}^{*-})&=&\frac{G_{F}}{\sqrt{2}}V_{ub}^{*}V_{cd}\left[a_{1}\mathcal
{A}_{ef}+C_{1}\mathcal {M}_{enf}\right],\\
\mathcal {A}(B_{s}^{0}\rightarrow
D_{s}^{+}K_{2}^{*-})&=&\frac{G_{F}}{\sqrt{2}}V_{ub}^{*}V_{cs}\left[a_{1}\mathcal
{A}_{ef}C_{1}\mathcal {M}_{enf}+a_{2}\mathcal {A}_{af}+C_{2}\mathcal
{M}_{anf}\right].
\end{eqnarray}

From Eq.(\ref{ffpmix}), we know that
\begin{eqnarray}
\mathcal {A}(B_{(s)}\rightarrow D^{(*)}f_{2})\,=\,\mathcal
{A}(B_{(s)}\rightarrow D^{(*)}f_{2}^{q})\cos\theta+\mathcal
{A}(B_{(s)}\rightarrow D^{(*)}f_{2}^{s})\sin\theta,\\
\mathcal {A}(B_{(s)}\rightarrow D^{(*)}f_{2}^{\prime})\,=\,\mathcal
{A}(B_{(s)}\rightarrow D^{(*)}f_{2}^{q})\sin\theta-\mathcal
{A}(B_{(s)}\rightarrow D^{(*)}f_{2}^{s})\cos\theta,
\end{eqnarray}
with $\theta=7.8^{\circ}$.

The diagrams for $\bar{b}\rightarrow \bar{c}$ decays are shown in
Fig.2 and Fig.3. The CKM favored decays are governed by the larger CKM matrix element
$V_{cb}$, then with a larger branching ratio.  Because a tensor
meson can not be produced through the (V $\pm$ A) or tensor current,
there are no factorizable emission diagrams with a tensor meson
emitted in Fig.3. We collect the decay amplitudes for each
$\bar{b}\rightarrow \bar{c}$ decays in Appendix \ref{b}.

\begin{figure}[]
\begin{center}
\vspace{-5cm} \centerline{\epsfxsize=10cm \epsffile{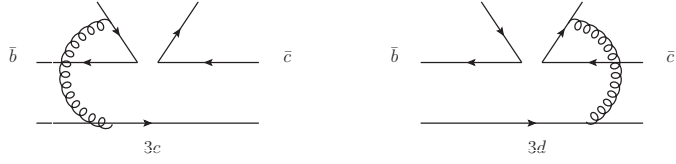}}
\vspace{-7cm} \caption{Feynman diagrams contributing to the
$B\,\rightarrow\,\bar{D}T$ decays with a tensor meson emitted}
 \label{fig:lodiagram3}
 \end{center}
\end{figure}

\section{NUMERICAL RESULTS AND DISCUSSIONS}

The decay width of a $B$ meson at rest decaying into $D(\bar{D})$
and $T$ is
\begin{eqnarray}
\Gamma(B\rightarrow
D(\bar{D})T)\,=\,\frac{|\overrightarrow{P}|}{8\pi
m_{B}^{2}}|\mathcal {A}(B\rightarrow D(\bar{D})T)|^{2},
\end{eqnarray}
where the momentum of the final state particle is given by
\begin{eqnarray}
|\overrightarrow
P|\,=\,\frac{1}{2m_{B}}\sqrt{\left[m_{B}^{2}-(m_{D}+m_{T})^{2}\right]
\left[m_{B}^{2}-(m_{D}-m_{T})^{2}\right]}.
\end{eqnarray}
For $B\rightarrow D^{*}(\bar{D}^{*})T$ decays, the decay width can
be written as
\begin{eqnarray}
\Gamma(B\rightarrow
D^{*}(\bar{D}^{*})T)\,=\,\frac{|\overrightarrow{P}|}{8\pi
m_{B}^{2}}\sum_{i=+,-,0}|\mathcal {A}_{i}(B\rightarrow
D^{*}(\bar{D}^{*})T)|^{2},
\end{eqnarray}
where the three polarization amplitudes $\mathcal {A}_{i}$ are given
by
\begin{eqnarray}
\mathcal {A}_{0}=\mathcal {A}^{L},\;\mathcal {A}_{\pm}=\mathcal
{A}^{T}\pm\mathcal {A}^{N}.
\end{eqnarray}

\begin{table}[]
\centering
 \caption{Branching ratios of $B_{(s)}\rightarrow DT$ decays calculated in the PQCD approach
 together with results from Isgur-Scora-Grinstein-Wise (ISGW) II model \cite{arxiv1010.3077,prd67014011} (unit:$10^{-7}$).}
 \vspace{0.2cm}
\begin{tabular}[t]{l!{\;\;\;\;}c!{\;\;\;\;}c!{\;\;\;\;}c!{\;\;\;\;}c}
\hline\hline

 \multirow{2}{*}{Decay Modes}& \multirow{2}{*}{Class} & \multirow{2}{*}{This Work} &
 \multirow{2}{*}{SDV[14]}
 & \multirow{2}{*}{KLO[15]}\\
 &&&&\\
 \hline
\vspace{0.2cm}
\multirow{1}{*}{$B^{0}\rightarrow D^{0}a_{2}^{0}$}&\multirow{1}{*}{C} &\multirow{1}{*}{$0.55_{-0.20\,-0.15\,-0.08}^{+0.28\,+0.15\,+0.10}$} &\multirow{1}{*}{0.34}&\multirow{1}{*}{...}\\
\vspace{0.2cm}
\multirow{1}{*}{$B^{0}\rightarrow D^{0}f_{2}$}& \multirow{1}{*}{C}& \multirow{1}{*}{$2.05_{-0.71\,-0.29\,-0.24}^{+0.81\,+0.34\,+0.27}$} & \multirow{1}{*}{0.36} &\multirow{1}{*}{...}\\
\vspace{0.2cm}
\multirow{1}{*}{$B^{0}\rightarrow D^{0}f_{2}^{\prime}$} & \multirow{1}{*}{C} & \multirow{1}{*}{$0.038_{-0.013\,-0.006\,-0.005}^{+0.015\,+0.006\,+0.005}$} & \multirow{1}{*}{$0.0071$} & \multirow{1}{*}{...}\\
\vspace{0.2cm}
\multirow{1}{*}{$B^{0}\rightarrow D^{0}K_{2}^{*0}$} & \multirow{1}{*}{C} & \multirow{1}{*}{$41.8_{-14.1\,-7.04\,-5.36}^{+17.4\,+7.50\,+5.75}$} & \multirow{1}{*}{12}& \multirow{1}{*}{11}\\
\vspace{0.2cm}
\multirow{1}{*}{$B^{0}\rightarrow D^{+}a_{2}^{-}$} & \multirow{1}{*}{T} & \multirow{1}{*}{$15.2_{-6.31\,-2.62\,-1.80}^{+7.82\,+1.97\,+1.96}$} & \multirow{1}{*}{12} & \multirow{1}{*}{...}\\
\vspace{0.2cm}
\multirow{1}{*}{$B^{0}\rightarrow D_{s}^{+}a_{2}^{-}$} & \multirow{1}{*}{T} & \multirow{1}{*}{$521_{-189\,-60\,-65}^{+249\,+44\,+72}$} & \multirow{1}{*}{$380$} & \multirow{1}{*}{$180$}\\
\vspace{0.2cm}
\multirow{1}{*}{$B^{0}\rightarrow D_{s}^{+}K_{2}^{*-}$} & \multirow{1}{*}{E} & \multirow{1}{*}{$0.61_{-0.14\,-0.16\,-0.07}^{+0.15\,+0.12\,+0.08}$} & \multirow{1}{*}{...} & \multirow{1}{*}{...}\\
\vspace{0.2cm}
\multirow{1}{*}{$B^{+}\rightarrow D^{0}a_{2}^{+}$} & \multirow{1}{*}{C} & \multirow{1}{*}{$1.95_{-0.70\,-0.48\,-0.24}^{+0.81\,+0.41\,+0.24}$} & \multirow{1}{*}{0.73} & \multirow{1}{*}{...}\\
\vspace{0.2cm}
\multirow{1}{*}{$B^{+}\rightarrow D^{0}K_{2}^{*+}$} & \multirow{1}{*}{C} & \multirow{1}{*}{$37.3_{-12.4\,-8.32\,-4.67}^{+14.3\,+6.99\,+5.10}$} & \multirow{1}{*}{13} & \multirow{1}{*}{12}\\
\vspace{0.2cm}
\multirow{1}{*}{$B^{+}\rightarrow D^{+}a_{2}^{0}$} & \multirow{1}{*}{T} & \multirow{1}{*}{$9.40_{-3.39\,-1.62\,-1.12}^{+4.59\,+1.15\,+1.20}$} & \multirow{1}{*}{6.5} & \multirow{1}{*}{...}\\
\vspace{0.2cm}
\multirow{1}{*}{$B^{+}\rightarrow D^{+}f_{2}$} & \multirow{1}{*}{T} & \multirow{1}{*}{$12.9_{-5.31\,-1.42\,-1.50}^{+6.31\,+0.90\,+1.60}$} & \multirow{1}{*}{6.9} & \multirow{1}{*}{...}\\
\vspace{0.2cm}
\multirow{1}{*}{$B^{+}\rightarrow D^{+}f_{2}^{\prime}$} & \multirow{1}{*}{T} & \multirow{1}{*}{$0.24_{-0.09\,-0.02\,-0.03}^{+0.12\,+0.02\,+0.03}$} & \multirow{1}{*}{1.4} & \multirow{1}{*}{...}\\
\vspace{0.2cm}
\multirow{1}{*}{$B^{+}\rightarrow D^{+}K_{2}^{*0}$} & \multirow{1}{*}{A} & \multirow{1}{*}{$5.27_{-1.65\,-0.66\,-0.66}^{+1.78\,+0.69\,+0.72}$} & \multirow{1}{*}{...} & \multirow{1}{*}{...}\\
\vspace{0.2cm}
\multirow{1}{*}{$B^{+}\rightarrow D_{s}^{+}a_{2}^{0}$} & \multirow{1}{*}{T} & \multirow{1}{*}{$280_{-110\,-33\,-35}^{+134\,+23\,+38}$} & \multirow{1}{*}{$200$} & \multirow{1}{*}{94}\\
\vspace{0.2cm}
\multirow{1}{*}{$B^{+}\rightarrow D_{s}^{+}f_{2}$} & \multirow{1}{*}{T} & \multirow{1}{*}{$299_{-122\,-33\,-37}^{+149\,+26\,+41}$} & \multirow{1}{*}{$220$} & \multirow{1}{*}{$100$}\\
\vspace{0.2cm}
\multirow{1}{*}{$B^{+}\rightarrow D_{s}^{+}f_{2}^{\prime}$}& \multirow{1}{*}{T,A} & \multirow{1}{*}{$4.12_{-1.98\,-0.78\,-0.51}^{+1.69\,+1.62\,+0.57}$} & \multirow{1}{*}{4.3} & \multirow{1}{*}{1.2}\\
\vspace{0.2cm}
\multirow{1}{*}{$B^{+}\rightarrow D_{s}^{+}\bar{K}_{2}^{*0}$} & \multirow{1}{*}{A} & \multirow{1}{*}{$0.34_{-0.10\,-0.06\,-0.04}^{+0.12\,+0.06\,+0.04}$} & \multirow{1}{*}{...} & \multirow{1}{*}{...}\\
\vspace{0.2cm}
\multirow{1}{*}{$B_{s}\rightarrow D^{0}a_{2}^{0}$} & \multirow{1}{*}{E} & \multirow{1}{*}{$3.87_{-1.19\,-0.95\,-0.48}^{+1.35\,+0.69\,+0.53}$} & \multirow{1}{*}{...} & \multirow{1}{*}{...}\\
\vspace{0.2cm}
\multirow{1}{*}{$B_{s}\rightarrow D^{0}f_{2}$} & \multirow{1}{*}{E} & \multirow{1}{*}{$6.26_{-1.99\,-1.18\,-0.48}^{+2.29\,+1.00\,+0.53}$} & \multirow{1}{*}{0.15} & \multirow{1}{*}{...}\\
\vspace{0.2cm}
\multirow{1}{*}{$B_{s}\rightarrow D^{0}f_{2}^{\prime}$} & \multirow{1}{*}{C} & \multirow{1}{*}{$25.5_{-11.4\,-3.35\,-3.2}^{+12.5\,+4.00\,+3.5}$} & \multirow{1}{*}{10}&\multirow{1}{*}{...}\\
\vspace{0.2cm}
\multirow{1}{*}{$B_{s}\rightarrow D^{0}\bar{K}_{2}^{*0}$} & \multirow{1}{*}{C}&\multirow{1}{*}{$1.42_{-0.55\,-0.19\,-0.17}^{+0.69\,+0.24\,+0.18}$} & \multirow{1}{*}{0.46} & \multirow{1}{*}{...}\\
\vspace{0.2cm}
\multirow{1}{*}{$B_{s}\rightarrow D^{+}a_{2}^{-}$} & \multirow{1}{*}{E} & \multirow{1}{*}{$8.06_{-2.68\,-1.99\,-1.00}^{+3.03\,+1.43\,+1.11}$} & \multirow{1}{*}{...} & \multirow{1}{*}{...}\\
\vspace{0.2cm}
\multirow{1}{*}{$B_{s}\rightarrow D^{+}K_{2}^{*-}$} & \multirow{1}{*}{T} & \multirow{1}{*}{$11.2_{-4.51\,-0.71\,-1.33}^{+5.61\,+0.60\,+1.46}$} & \multirow{1}{*}{8.3} & \multirow{1}{*}{...}\\
\vspace{0.4cm}
\multirow{1}{*}{$B_{s}\rightarrow D_{s}^{+}K_{2}^{*-}$}&\multirow{1}{*}{T} & \multirow{1}{*}{$206_{-95\,-26\,-25}^{+115\,+16\,+28}$} & \multirow{1}{*}{$260$} & \multirow{1}{*}{...}\\
 \hline\hline
\end{tabular}\label{s1}
\end{table}

\begin{table}[htb]
\centering
 \caption{Branching ratios (unit:$10^{-7}$) and the percentage of transverse polarizations $R_{T}$(unit:$\%$) of $B_{(s)}\rightarrow D^{*}T$ decays calculated in the PQCD approach
 together with results from ISGW II model \cite{arxiv1010.3077,prd67014011}.}
 \vspace{0.2cm}
\begin{tabular}[t]{lccccc}
\hline\hline
 \vspace{0.3cm}
 \multirow{3}{*}{Decay Modes}& \multirow{3}{*}{Class} &
 \multicolumn{3}{c}{ \multirow{1}{*}{Branching Ratio}}& \multirow{3}{*}{$R_{T}$}\\
\cline{3-5}
&&\multirow{2}{*}{This Work}&\multirow{2}{*}{SDV[14]}& \multirow{2}{*}{KLO[15]}&\\
&&&&&\\
 \hline
\vspace{0.18cm}
\multirow{1}{*}{$B^{0}\rightarrow D^{*0}a_{2}^{0}$}&\multirow{1}{*}{C} &
\multirow{1}{*}{$1.34_{-0.53\,-0.19\,-0.15}^{+0.64\,+0.25\,+0.17}$} &\multirow{1}{*}{0.50}&
\multirow{1}{*}{...}& \multirow{1}{*}{$47_{-4.5\,-1.6}^{+3.7\,+1.4}$}\\
\vspace{0.18cm}
\multirow{1}{*}{$B^{0}\rightarrow D^{*0}f_{2}$}& \multirow{1}{*}{C}&
\multirow{1}{*}{$2.70_{-1.02\,-0.30\,-0.33}^{+1.22\,+0.43\,+0.36}$} & \multirow{1}{*}{0.53} &
\multirow{1}{*}{...}& \multirow{1}{*}{$26_{-4.0\,-1.1}^{+3.7\,+1.3}$}\\
\vspace{0.18cm}
\multirow{1}{*}{$B^{0}\rightarrow D^{*0}f_{2}^{\prime}$} & \multirow{1}{*}{C} &
 \multirow{1}{*}{$0.052_{-0.02\,-0.005\,-0.006}^{+0.023\,+0.008\,+0.007}$} &
 \multirow{1}{*}{0.01} & \multirow{1}{*}{...}& \multirow{1}{*}{$26_{-4.0\,-1.1}^{+3.7\,+1.3}$}\\
\vspace{0.18cm}
\multirow{1}{*}{$B^{0}\rightarrow D^{*0}K_{2}^{*0}$} & \multirow{1}{*}{C} &
\multirow{1}{*}{$60.5_{-21.3\,-9.15\,-7.56}^{+25.3\,+10.6\,+8.30}$} & \multirow{1}{*}{19}&
\multirow{1}{*}{18}& \multirow{1}{*}{$22_{-3.2\,-0.5}^{+2.7\,+1.0}$}\\
\vspace{0.18cm}
\multirow{1}{*}{$B^{0}\rightarrow D^{*+}a_{2}^{-}$} & \multirow{1}{*}{T} &
\multirow{1}{*}{$21.6_{-8.60\,-3.10\,-2.50}^{+10.6\,+2.48\,+2.90}$} & \multirow{1}{*}{18} &
\multirow{1}{*}{...}&\multirow{1}{*}{$28_{-1.5\,-1.6}^{+1.6\,+1.7}$}\\
\vspace{0.18cm}
\multirow{1}{*}{$B^{0}\rightarrow D_{s}^{*+}a_{2}^{-}$} & \multirow{1}{*}{T} &
\multirow{1}{*}{$688_{-267\,-79\,-86}^{+321\,+55\,+94}$} &
\multirow{1}{*}{$367$} &
 \multirow{1}{*}{$291$}& \multirow{1}{*}{$26_{-0.9\,-0.7}^{+1.2\,+0.4}$}\\
\vspace{0.18cm}
\multirow{1}{*}{$B^{0}\rightarrow D_{s}^{*+}K_{2}^{*-}$} & \multirow{1}{*}{E} &
\multirow{1}{*}{$0.57_{-0.13\,-0.11\,-0.07}^{+0.13\,+0.12\,+0.07}$} & \multirow{1}{*}{...} &
\multirow{1}{*}{...}&\multirow{1}{*}{$12_{-1.6\,-2.5}^{+1.6\,+3.2}$}\\
\vspace{0.18cm}
\multirow{1}{*}{$B^{+}\rightarrow D^{*0}a_{2}^{+}$} & \multirow{1}{*}{C} &
\multirow{1}{*}{$4.46_{-1.65\,-0.61\,-0.53}^{+2.01\,+0.73\,+0.58}$} & \multirow{1}{*}{1.1} &
 \multirow{1}{*}{...}&\multirow{1}{*}{$41_{-3.6\,-1.6}^{+3.8\,+2.3}$}\\
\vspace{0.18cm}
\multirow{1}{*}{$B^{+}\rightarrow D^{*0}K_{2}^{*+}$} & \multirow{1}{*}{C} &
 \multirow{1}{*}{$72.1_{-23.7\,-9.38\,-9.00}^{+28.3\,+11.8\,+10.}$} & \multirow{1}{*}{21} &
  \multirow{1}{*}{19}&\multirow{1}{*}{$35_{-3.6\,-1.0}^{+4.0\,+0.9}$}\\
\vspace{0.18cm}
\multirow{1}{*}{$B^{+}\rightarrow D^{*+}a_{2}^{0}$} & \multirow{1}{*}{T} &
\multirow{1}{*}{$14.0_{-5.45\,-1.34\,-1.65}^{+6.51\,+1.03\,+1.80}$} &
\multirow{1}{*}{9.6} & \multirow{1}{*}{...}&\multirow{1}{*}{$25_{-1.0\,-0.1}^{+1.5\,+0.3}$}\\
\vspace{0.18cm}
\multirow{1}{*}{$B^{+}\rightarrow D^{*+}f_{2}$} & \multirow{1}{*}{T} &
\multirow{1}{*}{$15.1_{-6.27\,-2.15\,-1.90}^{+8.83\,+1.42\,+2.01}$} & \multirow{1}{*}{10} & \multirow{1}{*}{...}&
\multirow{1}{*}{$25_{-1.0\,-1.8}^{+1.3\,+1.8}$}\\
\vspace{0.18cm}
\multirow{1}{*}{$B^{+}\rightarrow D^{*+}f_{2}^{\prime}$} & \multirow{1}{*}{T} &
 \multirow{1}{*}{$0.29_{-0.11\,-0.04\,-0.02}^{+0.15\,+0.03\,+0.02}$} & \multirow{1}{*}{0.21} & \multirow{1}{*}{...}&\multirow{1}{*}{$25_{-1.0\,-1.8}^{+1.3\,+1.8}$}\\
\vspace{0.18cm}
\multirow{1}{*}{$B^{+}\rightarrow D^{*+}K_{2}^{*0}$} & \multirow{1}{*}{A} &
\multirow{1}{*}{$18.2_{-5.15\,-2.15\,-2.70}^{+4.77\,+0.21\,+2.00}$} & \multirow{1}{*}{...} &
 \multirow{1}{*}{...}& \multirow{1}{*}{$82_{-2.9\,-2.7}^{+2.1\,+3.8}$}\\
\vspace{0.18cm}
\multirow{1}{*}{$B^{+}\rightarrow D_{s}^{*+}a_{2}^{0}$} & \multirow{1}{*}{T} &
 \multirow{1}{*}{$330_{-127\,-37\,-42}^{+155\,+27\,+45}$} &
 \multirow{1}{*}{$196$} & \multirow{1}{*}{$155$}&\multirow{1}{*}{$26_{-0.8\,-0.7}^{+1.2\,+0.4}$}\\
\vspace{0.18cm}
\multirow{1}{*}{$B^{+}\rightarrow D_{s}^{*+}f_{2}$} & \multirow{1}{*}{T} &
\multirow{1}{*}{$385_{-156\,-44\,-48}^{+203\,+31\,+52}$} &
\multirow{1}{*}{$207$} & \multirow{1}{*}{$167$}&\multirow{1}{*}{$25_{-0.8\,-0.8}^{+1.1\,+0.7}$}\\
\vspace{0.18cm}
\multirow{1}{*}{$B^{+}\rightarrow D_{s}^{*+}f_{2}^{\prime}$}& \multirow{1}{*}{A} &
 \multirow{1}{*}{$21.6_{-6.03\,-2.32\,-2.70}^{+6.77\,+1.00\,+3.00}$} & \multirow{1}{*}{4.0} &
 \multirow{1}{*}{2.0}& \multirow{1}{*}{$83_{-5.3\,-1.9}^{+5.2\,+1.9}$}\\
\vspace{0.18cm}
\multirow{1}{*}{$B^{+}\rightarrow D_{s}^{*+}\bar{K}_{2}^{*0}$} & \multirow{1}{*}{A} &
\multirow{1}{*}{$1.25_{-0.34\,-0.16\,-0.15}^{+0.36\,+0.06\,+0.16}$} & \multirow{1}{*}{...} & \multirow{1}{*}{...}&\multirow{1}{*}{$81_{-1.8\,-3.3}^{+1.6\,+3.7}$}\\
\vspace{0.18cm}
\multirow{1}{*}{$B_{s}\rightarrow D^{*0}a_{2}^{0}$} & \multirow{1}{*}{E} &
\multirow{1}{*}{$2.68_{-0.81\,-0.63\,-0.33}^{+0.91\,+0.70\,+0.37}$} & \multirow{1}{*}{...} &
 \multirow{1}{*}{...}&\multirow{1}{*}{$21_{-3.0\,-3.9}^{+2.6\,+4.8}$}\\
\vspace{0.18cm}
\multirow{1}{*}{$B_{s}\rightarrow D^{*0}f_{2}$} & \multirow{1}{*}{E} &
\multirow{1}{*}{$5.06_{-1.65\,-0.98\,-0.62}^{+1.93\,+0.84\,+0.71}$} & \multirow{1}{*}{0.24} & \multirow{1}{*}{...}&\multirow{1}{*}{$14_{-2.1\,-1.6}^{+2.0\,+2.0}$}\\
\vspace{0.18cm}
\multirow{1}{*}{$B_{s}\rightarrow D^{*0}f_{2}^{\prime}$} & \multirow{1}{*}{C} &
\multirow{1}{*}{$36.2_{-14.4\,-5.51\,-4.60}^{+17.3\,+5.62\,+4.90}$} & \multirow{1}{*}{16}&\multirow{1}{*}{...}&\multirow{1}{*}{$17_{-3.1\,-1.1}^{+2.9\,+1.6}$}\\
\vspace{0.18cm}
\multirow{1}{*}{$B_{s}\rightarrow D^{*0}\bar{K}_{2}^{*0}$} & \multirow{1}{*}{C}&
\multirow{1}{*}{$2.06_{-0.83\,-0.31\,-0.24}^{+1.03\,+0.33\,+0.25}$} & \multirow{1}{*}{0.7} &
\multirow{1}{*}{...}&\multirow{1}{*}{$21_{-3.6\,-0.6}^{+3.0\,+0.8}$}\\
\vspace{0.18cm}
\multirow{1}{*}{$B_{s}\rightarrow D^{*+}a_{2}^{-}$} & \multirow{1}{*}{E} & \multirow{1}{*}{$5.36_{-1.59\,-1.27\,-0.66}^{+1.82\,+1.41\,+0.74}$} & \multirow{1}{*}{...} & \multirow{1}{*}{...}&\multirow{1}{*}{$21_{-3.0\,-3.9}^{+2.6\,+4.8}$}\\
\vspace{0.18cm}
\multirow{1}{*}{$B_{s}\rightarrow D^{*+}K_{2}^{*-}$} & \multirow{1}{*}{T} & \multirow{1}{*}{$14.8_{-5.93\,-0.85\,-1.77}^{+7.42\,+0.90\,+1.94}$} & \multirow{1}{*}{12} & \multirow{1}{*}{...}& \multirow{1}{*}{$26_{-1.0\,-0.2}^{+1.2\,-0.1}$}\\
\vspace{0.35cm}
\multirow{1}{*}{$B_{s}\rightarrow D_{s}^{*+}K_{2}^{*-}$} & \multirow{1}{*}{T} & \multirow{1}{*}{$332_{-138\,-31\,-41}^{+172\,+20\,+46}$} & \multirow{1}{*}{$261$} & \multirow{1}{*}{...}& \multirow{1}{*}{$34_{-1.6\,-1.0}^{+1.9\,+1.5}$}\\
 \hline\hline
\end{tabular}\label{s2}
\end{table}

\begin{table}[htb]
\centering
 \caption{Branching ratios of $B_{(s)}\rightarrow \bar{D}T$ decays calculated in the PQCD approach
 together with results from ISGW II model \cite{arxiv1010.3077,prd67014011} (unit:$10^{-5}$).}
 \vspace{0.2cm}
\begin{tabular}[t]{lcccc}
\hline\hline

 \multirow{2}{*}{Decay Modes}& \multirow{2}{*}{Class} & \multirow{2}{*}{This Work} &
 \multirow{2}{*}{SDV[14]}
 & \multirow{2}{*}{KLO[15]}\\
 &&&&\\
 \hline
\vspace{0.18cm}
\multirow{1}{*}{$B^{0}\rightarrow \bar{D}^{0}a_{2}^{0}$}&\multirow{1}{*}{C} &
\multirow{1}{*}{$12.3_{-2.99\,-3.24\,-0.40}^{+3.21\,+3.08\,+0.67}$} &\multirow{1}{*}{8.2}&\multirow{1}{*}{4.8}\\
\vspace{0.18cm}
\multirow{1}{*}{$B^{0}\rightarrow \bar{D}^{0}f_{2}$}& \multirow{1}{*}{C}&
\multirow{1}{*}{$9.46_{-2.29\,-3.73\,-0.32}^{+2.52\,+3.64\,+0.51}$} & \multirow{1}{*}{8.8} & \multirow{1}{*}{5.3}\\
\vspace{0.18cm}
\multirow{1}{*}{$B^{0}\rightarrow \bar{D}^{0}f_{2}^{\prime}$} & \multirow{1}{*}{C} &
\multirow{1}{*}{$0.18_{-0.05\,-0.07\,-0.06}^{+0.04\,+0.06\,-0.09}$} & \multirow{1}{*}{0.17} & \multirow{1}{*}{0.062}\\
\vspace{0.18cm}
\multirow{1}{*}{$B^{0}\rightarrow \bar{D}^{0}K_{2}^{*0}$} & \multirow{1}{*}{C} &
\multirow{1}{*}{$1.45_{-0.38\,-0.33\,-0.05}^{+0.41\,+0.29\,-0.09}$} & \multirow{1}{*}{0.81}& \multirow{1}{*}{0.68}\\
\vspace{0.18cm}
\multirow{1}{*}{$B^{0}\rightarrow D^{-}a_{2}^{+}$} & \multirow{1}{*}{T} &
 \multirow{1}{*}{$39.8_{-12.6\,-12.1\,-1.34}^{+15.3\,+12.5\,+2.15}$} & \multirow{1}{*}{...} & \multirow{1}{*}{...}\\
\vspace{0.18cm}
\multirow{1}{*}{$B^{0}\rightarrow D^{-}K_{2}^{*+}$} & \multirow{1}{*}{T} &
\multirow{1}{*}{$1.16_{-0.40\,-0.47\,-0.05}^{+0.50\,+0.52\,+0.06}$} & \multirow{1}{*}{...} & \multirow{1}{*}{...}\\
\vspace{0.18cm}
\multirow{1}{*}{$B^{0}\rightarrow D_{s}^{-}K_{2}^{*+}$} & \multirow{1}{*}{E} &
\multirow{1}{*}{$6.06_{-1.65\,-1.04\,-0.21}^{+1.73\,+0.43\,+0.32}$} & \multirow{1}{*}{...} & \multirow{1}{*}{...}\\
\vspace{0.18cm}
\multirow{1}{*}{$B^{+}\rightarrow \bar{D}^{0}a_{2}^{+}$} & \multirow{1}{*}{T,C} &
\multirow{1}{*}{$41.5_{-12.6\,-14.2\,-1.40}^{+16.5\,+13.0\,+2.24}$} & \multirow{1}{*}{18} & \multirow{1}{*}{10}\\
\vspace{0.18cm}
\multirow{1}{*}{$B^{+}\rightarrow \bar{D}^{0}K_{2}^{*+}$} & \multirow{1}{*}{T,C} &
\multirow{1}{*}{$3.33_{-1.02\,-0.91\,-0.13}^{+1.33\,+0.87\,+0.20}$} & \multirow{1}{*}{0.87} & \multirow{1}{*}{0.73}\\
\vspace{0.18cm}
\multirow{1}{*}{$B_{s}\rightarrow \bar{D}^{0}a_{2}^{0}$} & \multirow{1}{*}{E} &
\multirow{1}{*}{$0.11_{-0.04\,-0.02\,-0.01}^{+0.04\,+0.01\,+0.01}$} & \multirow{1}{*}{...} & \multirow{1}{*}{...}\\
\vspace{0.18cm}
\multirow{1}{*}{$B_{s}\rightarrow \bar{D}^{0}f_{2}$} & \multirow{1}{*}{E} &
\multirow{1}{*}{$0.14_{-0.05\,-0.03\,-0.01}^{+0.04\,+0.01\,+0.01}$} & \multirow{1}{*}{$0.0099$} &
\multirow{1}{*}{...}\\
\vspace{0.18cm}
\multirow{1}{*}{$B_{s}\rightarrow \bar{D}^{0}f_{2}^{\prime}$} & \multirow{1}{*}{C} &
\multirow{1}{*}{$1.36_{-0.43\,-0.26\,-0.06}^{+0.53\,+0.22\,+0.08}$} & \multirow{1}{*}{0.67}&\multirow{1}{*}{...}\\
\vspace{0.18cm}
\multirow{1}{*}{$B_{s}\rightarrow \bar{D}^{0}\bar{K}_{2}^{*0}$} & \multirow{1}{*}{C} &
 \multirow{1}{*}{$20.3_{-6.41\,-4.59\,-0.80}^{+7.70\,+3.98\,+1.00}$} & \multirow{1}{*}{11} & \multirow{1}{*}{...}\\
\vspace{0.18cm}
\multirow{1}{*}{$B_{s}\rightarrow D^{-}a_{2}^{+}$} & \multirow{1}{*}{E} &
\multirow{1}{*}{$0.23_{-0.08\,-0.04\,-0.01}^{+0.08\,+0.02\,+0.01}$} & \multirow{1}{*}{...} & \multirow{1}{*}{...}\\
\vspace{0.18cm}
\multirow{1}{*}{$B_{s}\rightarrow D_{s}^{-}a_{2}^{+}$} & \multirow{1}{*}{T} &
\multirow{1}{*}{$11.3_{-4.43\,-4.88\,-0.38}^{+5.53\,+6.11\,+0.61}$} & \multirow{1}{*}{...} & \multirow{1}{*}{...}\\
\vspace{0.35cm}
\multirow{1}{*}{$B_{s}\rightarrow D_{s}^{-}K_{2}^{*+}$}&\multirow{1}{*}{T,E} &
\multirow{1}{*}{$1.97_{-0.69\,-0.67\,-0.08}^{+0.81\,+0.72\,+0.12}$} & \multirow{1}{*}{...} & \multirow{1}{*}{...}\\
 \hline\hline
\end{tabular}\label{s3}
\end{table}

\begin{table}[htb]
\centering
 \caption{Branching ratios (unit:$10^{-5}$) and the percentage of transverse polarizations $R_{T}$(unit:$\%$) of
 $B_{(s)}\rightarrow \bar{D}^{*}T$ decays calculated in PQCD approach
 together with results from ISGW II model \cite{arxiv1010.3077,prd67014011}.}
 \vspace{0.2cm}
\begin{tabular}[t]{lccccc}
\hline\hline
 \vspace{0.3cm}
 \multirow{3}{*}{Decay Modes}& \multirow{3}{*}{Class} & \multicolumn{3}{c}{ \multirow{1}{*}{Branching Ratio}}&
  \multirow{3}{*}{$R_{T}$}\\
\cline{3-5}
&&\multirow{2}{*}{This Work}&\multirow{2}{*}{SDV[14]}& \multirow{2}{*}{KLO[15]}&\\
&&&&&\\
\hline \vspace{0.18cm}
\multirow{1}{*}{$B^{0}\rightarrow \bar{D}^{*0}a_{2}^{0}$}&\multirow{1}{*}{C} &
\multirow{1}{*}{$39.3_{-11.1\,-0.50\,-1.34}^{+13.6\,+2.05\,+2.15}$} &\multirow{1}{*}{12}&
\multirow{1}{*}{7.8}& \multirow{1}{*}{$73_{-4.3\,-8.1}^{+4.6\,+9.0}$}\\
\vspace{0.18cm}
\multirow{1}{*}{$B^{0}\rightarrow \bar{D}^{*0}f_{2}$}& \multirow{1}{*}{C}&
\multirow{1}{*}{$38.2_{-11.6\,-1.22\,-1.30}^{+13.9\,+1.97\,+2.10}$} & \multirow{1}{*}{13} &
\multirow{1}{*}{8.4}&\multirow{1}{*}{$70_{-5.6\,-6.3}^{+5.9\,+9.4}$}\\
\vspace{0.18cm}
\multirow{1}{*}{$B^{0}\rightarrow \bar{D}^{*0}f_{2}^{\prime}$} & \multirow{1}{*}{C} &
\multirow{1}{*}{$0.72_{-0.22\,-0.03\,-0.03}^{+0.26\,+0.02\,+0.04}$} & \multirow{1}{*}{0.26}
& \multirow{1}{*}{0.11}&\multirow{1}{*}{$70_{-5.6\,-6.3}^{+5.9\,+9.4}$}\\
\vspace{0.18cm}
\multirow{1}{*}{$B^{0}\rightarrow \bar{D}^{*0}K_{2}^{*0}$} & \multirow{1}{*}{C} &
\multirow{1}{*}{$5.32_{-1.42\,-0.66\,-0.22}^{+1.69\,+0.79\,+0.32}$} & \multirow{1}{*}{1.3}&
 \multirow{1}{*}{1.1}&\multirow{1}{*}{$71_{-1.6\,-8.8}^{+1.8\,+8.6}$}\\
\vspace{0.18cm}
\multirow{1}{*}{$B^{0}\rightarrow D^{*-}a_{2}^{+}$} & \multirow{1}{*}{T} &
\multirow{1}{*}{$29.6_{-9.86\,-10.1\,-1.01}^{+11.5\,+9.58\,+1.62}$} & \multirow{1}{*}{...} &
\multirow{1}{*}{...}&\multirow{1}{*}{$3_{-0.2\,-0.4}^{+0.2\,+0.4}$}\\
\vspace{0.18cm}
\multirow{1}{*}{$B^{0}\rightarrow D^{*-}K_{2}^{*+}$} & \multirow{1}{*}{T} &
\multirow{1}{*}{$1.15_{-0.37\,-0.43\,-0.04}^{+0.49\,+0.46\,+0.07}$} & \multirow{1}{*}{...} &
\multirow{1}{*}{...}&\multirow{1}{*}{$7_{-0.2\,-0.3}^{+0.4\,+0.3}$}\\
\vspace{0.18cm} \multirow{1}{*}{$B^{0}\rightarrow
D_{s}^{*-}K_{2}^{*+}$} & \multirow{1}{*}{E} & \
\multirow{1}{*}{$4.55_{-1.14\,-0.51\,-0.16}^{+1.32\,+0.48\,+0.25}$}
& \multirow{1}{*}{...} &
\multirow{1}{*}{...}&\multirow{1}{*}{$22_{-3.4\,-6.7}^{+3.0\,+7.6}$}\\
\vspace{0.18cm}
\multirow{1}{*}{$B^{+}\rightarrow \bar{D}^{*0}a_{2}^{+}$} & \multirow{1}{*}{T,C} &
\multirow{1}{*}{$80.6_{-24.1\,-3.12\,-2.75}^{+29.0\,+3.67\,+4.41}$} & \multirow{1}{*}{26} &
\multirow{1}{*}{17}&\multirow{1}{*}{$58_{-3.2\,-10.2}^{+3.7\,+13.8}$}\\
\vspace{0.18cm}
\multirow{1}{*}{$B^{+}\rightarrow \bar{D}^{*0}K_{2}^{*+}$} & \multirow{1}{*}{T,C} &
\multirow{1}{*}{$6.81_{-1.93\,-0.30\,-0.27}^{+2.36\,+0.34\,+0.42}$} & \multirow{1}{*}{1.4} &
\multirow{1}{*}{1.2}&\multirow{1}{*}{$57_{-1.3\,-11.6}^{+1.4\,+14.1}$}\\
\vspace{0.18cm}
\multirow{1}{*}{$B_{s}\rightarrow \bar{D}^{*0}a_{2}^{0}$} & \multirow{1}{*}{E} &
\multirow{1}{*}{$0.09_{-0.03\,-0.01\,-0.01}^{+0.03\,+0.01\,+0.01}$} & \multirow{1}{*}{...} &
\multirow{1}{*}{...}&\multirow{1}{*}{$26_{-4.4\,-6.8}^{+4.0\,+6.5}$}\\
\vspace{0.18cm}
\multirow{1}{*}{$B_{s}\rightarrow \bar{D}^{*0}f_{2}$} & \multirow{1}{*}{E} &
\multirow{1}{*}{$0.21_{-0.07\,-0.02\,-0.01}^{+0.08\,+0.01\,+0.01}$} & \multirow{1}{*}{$0.016$} &
 \multirow{1}{*}{...}&\multirow{1}{*}{$11_{-0.6\,-1.0}^{+0.6\,+2.2}$}\\
\vspace{0.18cm}
\multirow{1}{*}{$B_{s}\rightarrow \bar{D}^{*0}f_{2}^{\prime}$} & \multirow{1}{*}{C} &
\multirow{1}{*}{$5.02_{-1.70\,-0.50\,-0.20}^{+2.06\,+0.39\,+0.30}$} & \multirow{1}{*}{1.1}&
\multirow{1}{*}{...}&\multirow{1}{*}{$71_{-1.0\,-7.2}^{+1.1\,+7.5}$}\\
\vspace{0.18cm}
\multirow{1}{*}{$B_{s}\rightarrow \bar{D}^{*0}\bar{K}^{*0}$} & \multirow{1}{*}{C} &
 \multirow{1}{*}{$70.1_{-23.9\,-5.82\,-2.40}^{+28.8\,+4.10\,+3.83}$} & \multirow{1}{*}{17} & \multirow{1}{*}{...}&
 \multirow{1}{*}{$68_{-1.3\,-7.9}^{+1.7\,+8.5}$}\\
\vspace{0.18cm}
\multirow{1}{*}{$B_{s}\rightarrow D^{*-}a_{2}^{+}$} & \multirow{1}{*}{E} &
 \multirow{1}{*}{$0.18_{-0.06\,-0.03\,-0.01}^{+0.06\,+0.01\,+0.01}$} & \multirow{1}{*}{...} &
 \multirow{1}{*}{...}&\multirow{1}{*}{$26_{-4.4\,-6.8}^{+4.0\,+6.5}$}\\
\vspace{0.18cm}
\multirow{1}{*}{$B_{s}\rightarrow D_{s}^{*-}a_{2}^{+}$} & \multirow{1}{*}{T} &
\multirow{1}{*}{$11.5_{-4.44\,-4.70\,-0.40}^{+5.59\,+5.79\,+0.63}$} & \multirow{1}{*}{...} &
\multirow{1}{*}{...}&\multirow{1}{*}{$6_{-0.1\,-0.3}^{+0.1\,+0.2}$}\\
\vspace{0.35cm}
\multirow{1}{*}{$B_{s}\rightarrow D_{s}^{*-}K_{2}^{*+}$}&\multirow{1}{*}{T,E} &
\multirow{1}{*}{$1.73_{-0.60\,-0.58\,-0.06}^{+0.75\,+0.65\,+0.11}$} & \multirow{1}{*}{...} &
\multirow{1}{*}{...}&\multirow{1}{*}{$11_{-0.9\,-2.4}^{+0.9\,+2.3}$}\\
 \hline\hline
\end{tabular}\label{s4}
\end{table}

All the input parameters, such as decay constants, CKM elements are
given in Appendix A, if not given in the previous two sections. The numerical results of branching ratios   for
the considered decay modes
  are summarized in
Tables~\ref{s1}$-$~\ref{s4}. We also show the results from ISGW II
model \cite{arxiv1010.3077,prd67014011} in these tables for
comparison if applicable. For those decays with a tensor meson
emitted and most of the pure annihilation type decays, our results
are the first time theoretical predictions. For the theoretical
uncertainties, we estimate three kinds of them: The first errors are
caused by the hadronic parameters, such as the decay constants and
the shape parameters in wave functions of charmed meson and the
$B_{(s)}$ meson, which are given in Sec.~\ref{sec:function}, and the
decay constants of tensor mesons given in Appendix ~\ref{a}. The second
errors are estimated from the unknown next-to-leading order QCD
corrections with respect to $\alpha_{s}$ and nonperturbative power
corrections with respect to scales in Sudakov exponents,
characterized by the choice of the
$\Lambda_{QCD}\,=\,(0.25\,\pm\,0.05)$ GeV and the variations of the
factorization scales shown in Appendix \ref{a}. The third error is from
the uncertainties of the CKM matrix elements. It is easy to see that
the most important theoretical uncertainty is caused by the
non-perturbative hadronic parameters, which can be improved by later
experiments.

As we know that all these decays do not have contributions from the
penguin operators. There are only four types of topology diagrams:
the color allowed diagrams (T), the color suppressed diagrams (C),
the W annihilation diagrams (A) and the W exchange diagrams (E). All
 decays are thus classified in the tables according to their
dominant contribution. Compared with $B\rightarrow D^{(*)} T$
decays, the $B\rightarrow \bar{D}^{(*)}T$ decays are enhanced by the
CKM matrix elements $|V_{cb}/V_{ub}|^2$, especially for those
without a strange quark in the four-quark operators. So for most of
the $B\rightarrow D^{(*)} T$ decays, the branching ratios are at the
order $10^{-6}$ or $10^{-7}$; while for the $B\rightarrow
\bar{D}^{(*)}T$ decays, the branching ratios are at the order
$10^{-4}$ or $10^{-5}$.

As usual, the nonfactorizable emission diagrams with a light meson
emitted are suppressed, because the contributions from two diagrams
cancel with each other. However, when the emitted meson is the
$D(\bar{D})$ or tensor meson, the situation is changed. Unlike the
light meson, the difference between $c(\bar{c})$ quark and the light
quark is very big in the heavy $D(\bar{D})$ meson
\cite{liprd78014018,zouhaojpg37}. The nonfactorizable diagrams also
provide non-negligible contributions. When the tensor meson is
emitted, the contributions from two nonfactorizable diagrams shown
in Fig.3 strengthen with each other, because the wave function of
tensor meson is antisymmetric under the interchange of the momentum
fractions of the quark and antiquark \cite{zheng1,zheng2}. Since the
factorizable emission diagrams with a tensor meson emitted are
prohibited, the contribution of nonfacotrizbale emission diagrams
play the decisive role. For these color suppressed decay channels,
since the facotrizable contribution is suppressed by the Wilson
coefficient $a_{2}$ $(C_{1}+C_{2}/3)\simeq0.1$, while the Wilson coefficient
for nonfactorizable contribution is $C_{2}\simeq1.0$, the nonfactorizable
contribution plays the crucial role in the amplitude.  From table I to IV, one can see that
for the color suppressed decay modes, the predicted branching ratios
in the PQCD approach are larger than those of
Ref.\cite{arxiv1010.3077} and Ref.\cite{prd67014011}. For
$B^{0}\rightarrow \bar{D}^{0}f_{2}$, our predicted branching ratio
$\mathcal {B}(B^{0}\rightarrow
\bar{D}^{0}f_{2})\,=\,9.46\times10^{-5}$, which is larger than other
approaches,  agrees better with the experimental data
$(12\pm4)\times10^{-5}$ \cite{jpg37075021}. In addition,
the annihilation diagrams can also provide relatively sizable
contributions. Our results show that the contributions from
annihilation diagrams are even at the same order as the emission
diagrams in some decay modes. Some of the pure annihilation type decays are already discussed in Ref. \cite{zou-zhou} with large branching ratios.

For those color allowed decay channels, the Wilson coefficient for
factorizable contribution is $a_{1}=(C_{1}/3+C_{2})\simeq 1$, while
for the nonfactorizable contribution is $C_{1}\simeq -0.3$. The
contribution of nonfactorizable diagrams is highly suppressed by the
Wilson coefficient. The decay amplitude is dominated by the
contribution from factorizable emission diagrams, which can be
naively factorized as the product of the Wilson coefficient $a_{1}$,
the decay constant of $D$ meson and the $B$ to tensor meson form
factor. In this case, our predicted branching ratios basically agree
with the predictions of naive factorization approach in
Ref.\cite{arxiv1010.3077}. The small difference is caused by
parameter changes and the interference from nonfactorizable and
annihilation diagrams. For those decays with a tensor meson emitted,
for example, $B^{0}\rightarrow D^{-}a_{2}^{+}$, since the
factorizable emission diagrams are prohibited, the predictions can
not be given within the naive factorization framework. But these
decays can get contributions from nonfatorizable and annihilation
type diagrams, which can be calculated in the PQCD approach.   The
branching ratios of these decays are predicted for the first time in
table \ref{s1}-\ref{s4}.

Similar to the relation $\mathcal {B}(B^{0}\rightarrow
\bar{D}^{(*)0}\rho^{0})\,>\,\mathcal {B}(B^{0}\rightarrow
\bar{D}^{(*)0}\omega)$ \cite{fd1}, we also get $\mathcal
{B}(B^{0}\rightarrow \bar{D}^{(*)0}a_{2}^{0})\,>\,\mathcal
{B}(B^{0}\rightarrow \bar{D}^{(*)0}f_{2})$. This can be explained by
the interference between contributions from emission diagram (C) and
contributions from annihilation diagrams (E). The interference can
also explain why $\mathcal {B}(B^{+}\rightarrow
D^{(*)+}f_{2})\,>\,\mathcal {B}(B^{+}\rightarrow
D^{(*)+}a_{2}^{0})$. The relative sign of the annihilation diagrams
(A) with respect to the emission diagrams (T) is negative for the
$a_{2}$ meson and positive for $f_{2}$ meson. The interference is constructive
for $B^{+}\rightarrow D^{(*)+}f_{2}$, while destructive for
$B^{+}\rightarrow D^{(*)+}a_{2}^{0}$.

For   decays involving $f_{2}^{(\prime)}$ in the final states, for
example, $B^{0}\rightarrow D^{0}f_{2}^{(\prime)}$, there are no
contributions from $s\bar s$ component. The branching ratios have
the simple relation derived from Eq.(\ref{ffpmix}):
\begin{eqnarray}
r\,=\,\frac{\mathcal {B}(B\rightarrow D f_{2}^{\prime})}{\mathcal
{B}(B\rightarrow D
f_{2})}\,=\,\frac{\sin^{2}\theta}{\cos^{2}\theta}.
\end{eqnarray}
This provides a potential way to measure the mixing angle of $f_2$
and $f_2'$, for example, $r \simeq\,0.02  $ with
$\theta=7.8^{\circ}$.

For $B\rightarrow D^{*}(\bar{D}^{*}) T$ decays, we also calculate
the percentage of transverse polarizations
\begin{eqnarray}
R_{T}\,=\,\frac{\mathcal {A}_{+}^{2}+\mathcal {A}_{-}^{2}}{\mathcal
{A}_{0}^{2}+\mathcal {A}_{+}^{2}+\mathcal {A}_{-}^{2}}.
\end{eqnarray}
The numerical results shown in table \ref{s2} and \ref{s4} are only
indicative, because the transversely polarized contributions are
suppressed by $r_{T}$ or $r_{D^*}$ to make it more sensitive to
meson wave function parameters and higher order corrections
\cite{liprd78014018}. According to the power counting rules in the
factorization assumption, the longitudinal polarization should be
dominant due to the quark helicity analysis
\cite{helicity1,helicity2}. This is true for those color favored
decay channels, such as $B^0 \to D^{*-} a_2^+$, $B^0 \to D^{*-}
K_2^{*+}$, $B_s^0 \to D_s^{*-} a_2^+$, $B_s^0 \to D_s^{*-}
K_2^{*+}$.

However, for those color suppressed (C) $B\rightarrow \bar{D}^{*}T$
($\bar{b}\rightarrow \bar{c}$ transition) decays with the
$\bar{D}^{*}$ emitted, the percentage of transverse polarizations
are about $70\%$, while for color suppressed (C) $B\rightarrow
{D}^{*}T$ ($\bar{b}\rightarrow \bar{u}$ transition) decays with
$D^{*}$ meson emitted, the percentage of transverse polarizations
are only at the range of $20\%$ to $30\%$.  For $B\rightarrow
\bar{D}^{*}T$ decays, we know that the $\bar{c}$ quark and the $u$
quark in $\bar{D}^{*}$ meson produced through $(V-A)$ current are
right-handed and left-handed respectively. So the $\bar{D}^{*}$
meson is longitudinally polarized. But the helicity of $\bar c$
quark can flip easily from right handed to left handed, because the
$\bar c$ quark is massive. Therefore the $\bar{D}^{*}$ can be
transversely polarized with the polarization $\lambda=-1$. The
recoiled tensor meson can also be transversely polarized with
polarization $\lambda=-1$ due to the contribution of orbital angular
momentum. Thus, the transversely polarized contribution can be
sizable. For $B\rightarrow {D}^{*}T$ decays, the emitted $D^{*}$
meson can also be transversely polarized, but the polarization is
$\lambda=+1$. The reason is that the $\bar u$ quark in ${D}^{*}$
meson is right handed, while the $c$ quark can flip from left handed
to right handed to make a $D^{*}$ meson with $\lambda=+1$.  The
recoiled transversely polarized tensor meson with polarization
$\lambda=+1$ needs contributions from both orbital angular momentum
and spin, so the situation is symmetric. But the wave function of
tensor meson is asymmetric. Therefore the transversely polarized
contribution is suppressed, because of
 Bose statistics.

As discussed in Ref.\cite{zouhaojpg37}, the W annihilation diagrams
give a very large contribution of transverse polarizations. In our
calculations, we also find    very large transverse polarizations up
to 80\% for the W annihilation (A) type  $B\rightarrow D^{*}T$
decays, such as $B^+ \to D^{*+} K_2^{*0}$, $B^+ \to D_s^{*+} \bar
K_2^{*0}$ and $B^+ \to D_s^{*+} f_2'$ decays. This can be understood
as following \cite{polarization}: For the $D^*$ meson, the ``light
quark-unti-quark" pair created from hard gluon are left-handed or
right-handed with equal opportunity. So the $D^{*}$ meson can be
longitudinally polarized or transversely polarized with polarization
$\lambda=-1$. For the tensor meson, the anti-quark from four quark
operator is right-handed, and the quark produced from hard gluon can
be either left-handed or right-handed. So the tensor meson can be
longitudinally polarized or transversely polarized with polarization
$\lambda=-1$, because of the additional contribution from the
orbital angular momentum. So the transverse polarization can become
so large with additional interference from other diagrams. Although
annihilation type diagrams, the W exchange diagrams (E) contribute
little to transverse polarizations, which is consistent with the
argument in $B\rightarrow D^{*}V$ decays
\cite{liprd78014018,zouhaojpg37}. Examples are $B_s \to D^{*0}
a_2^0$, $B_s \to \bar D^{*0} a_2^0$, $B_s \to D^{*0} f_2$, $B_s \to
\bar D^{*0} f_2$, $B_s \to D^{*+} a_2^-$, $B_s \to D^{*-} a_2^+$,
$B^0 \to D_s^{*+} K_2^{*-}$, $B^0 \to D_s^{*-} K_2^{*+}$, with only
10-20\% transverse polarization contributions.

\section{SUMMARY}

In this paper, we investigate the $B_{(s)}\rightarrow
D^{(*)}T,\bar{D}^{(*)}T$ decays within the framework of perturbative
QCD approach. We calculate the contributions of different
diagrams in the leading order approximation of $m_{D}/m_{B}$
expansion. We find that the nonfactorizable and annihilation type diagrams
provide large contributions, especially for those color suppressed
channels and the decays with a tensor meson emitted. We predict the
branching ratios and the ratios of the transverse polarized
contributions and find that the branching ratios for
$B_{(s)}\rightarrow D^{(*)}T$ decays are in the range of $10^{-5}$
to $10^{-8}$, while $10^{-4}$ to $10^{-6}$ for $B_{(s)}\rightarrow
\bar{D}^{(*)}T$ decays. For those color suppressed
$B_{(s)}\rightarrow \bar{D}^{(*)}T$ decays, the transversely
polarized contributions from nonfactorizable diagrams are very
large. For those W annihilation type $B\rightarrow D^{*}T$ decays,
the transverse polarized contributions from factorizable
annihilation diagrams are  as large as $80\%$.

\section*{Acknowledgment}

 We are very grateful to  Xin Liu and
 Run-Hui Li for helpful discussions. This Work is partly supported by
the National
Science Foundation of China under the Grant No.11075168.

\begin{appendix}

\section{Input Parameters and Hard Functions}\label{a}


\begin{table}[!htbh]
\centering
 \caption{The masses and decay constants of light tensor mesons \cite{TM,TK,zheng1}}
 \vspace{0.3cm}
\begin{tabular}{c!{\;\;\;\;\;\;}c!{\;\;\;\;\;\;}c}
\hline\hline
 \vspace{0.3cm}
 \multirow{1}{*}{Tensor(mass(MeV))} & \multirow{1}{*}{$f_{T}$(MeV)} &\multirow{1}{*}{$f_{T}^{\perp}$(MeV)} \\
\hline
 \vspace{0.1cm}
\multirow{1}{*}{$f_{2}(1270)$}&\multirow{1}{*}{$102\,\pm\,6$}&\multirow{1}{*}{$117\,\pm\,25$}\\
\vspace{0.1cm}
\multirow{1}{*}{$f_{2}^{\prime}(1525)$}&\multirow{1}{*}{$126\,\pm\,4$}&\multirow{1}{*}{$65\,\pm\,12$}\\
\vspace{0.1cm}
\multirow{1}{*}{$a_{2}(1320)$}&\multirow{1}{*}{$107\,\pm\,6$}&\multirow{1}{*}{$105\,\pm\,21$}\\
\vspace{0.3cm}
\multirow{1}{*}{$K_{2}^{*}(1430)$}&\multirow{1}{*}{$118\,\pm\,5$}&\multirow{1}{*}{$77\,\pm\,14$}\\
 \hline\hline
\end{tabular}\label{S}
\end{table}

The masses and decay constants of tensor mesons are summarized in
Table V. Other parameters such as QCD scale
$\Lambda_{\overline{MS}}^{f=4}=0.25$ GeV and the b quark mass
$m_{b}=4.8$(GeV). For the CKM matrix elements, here we adopt the
Wolfenstein parameterization   $A=0.808$, $\lambda=0.2253$,
$\bar{\rho}=0.132$ and $\bar{\eta}=0.341$ \cite{jpg37075021}.

Like the $\eta\,-\,\eta^{\prime}$ mixing, the iso-singlet tensor
meson states $f_{2}(1270)$ and $f_{2}^{\prime}(1525)$ are also  a
mixture of $f_{2}^{q}\,=\,\frac{1}{\sqrt{2}}(u\bar{u}\,+\,d\bar{d})$
and $f_{2}^{s}\,=\,s\bar{s}$
\begin{eqnarray}
&&f_{2}\,=\,f_{2}^{q}\cos\theta\,+\, f_{2}^{s}\sin\theta,\nonumber\\
&&f_{2}^{\prime}\,=\,f_{2}^{q}\sin\theta\,-\,f_{2}^{s}\cos\theta,
\label{ffpmix}
\end{eqnarray}
with the mixing angle $\theta\,=\,5.8^{\circ}$ \cite{zheng3},
$7.8^{\circ}$ \cite{jpg27807} or $(9\,\pm\,1)^{\circ}$
\cite{jpg37075021}.


  The functions $h$ in the decay amplitudes
consist of two parts: the jet function $S_{t}(x_{i})$ and the
propagator of virtual quark and gluon. The former is gained by the
threshold re-summation \cite{prd66094010}. For factorizable emission
diagrams Fig.1. 1a and 1b, the $h$ function can be given by
\begin{eqnarray}
h_{ef}(x_{1},x_{3},b_{1},b_{3})\,&=&\,K_{0}(\sqrt{x_{1}x_{3}}m_{B}b_{1})\nonumber\\
&&\times\left\{\theta{b_{1}-b_{3}}K_{0}\left(\sqrt{x_{3}}m_{B}b_{1}\right)I_{0}\left(\sqrt{x_{3}}m_{B}b_{3}\right)\right.
\nonumber\\
&&\left.+\theta(b_{3}-b_{1})K_{0}\left(\sqrt{x_{3}}m_{B}b_{3}\right)I_{0}\left(\sqrt{x_{3}}m_{B}b_{1}\right)\right\}\nonumber\\
&&\times S_{t}(x_{3}).
\end{eqnarray}
The hard scales are determined by
\begin{eqnarray}
&&t_{a}\,=\,\max\{\sqrt{x_{3}(1-r_{D}^{2})}m_{B},\,1/b_{1},\,1/b_{3}\},\nonumber\\
&&t_{b}\,=\,\max\{\sqrt{x_{1}(1-r_{D}^{2})}m_{B},\,1/b_{1},\,1/b_{3}\}.
\end{eqnarray}
Jet function appears in the factorization formulae is
\cite{prd66094010}
\begin{eqnarray}
S_{t}(x)\,=\,\frac{2^{1+2c}\Gamma(3/2+c)}{\sqrt{\pi}\Gamma(1+c)}[x(1-x)]^{c},
\end{eqnarray}
where $c\,=\,0.5$. For the nonfactorizable diagrams, we omit the
$S_{t}(x)$, because it provides a very small numerical effect to the
amplitude \cite{plb555}.

The evolution factors $E_{ef}(t_{a})$ and $E_{ef}(t_{b})$ in the
matrix elements (see section III) are given by
\begin{eqnarray}
E_{ef}(t)\,=\,\alpha_{s}(t)\exp[-S_{B}(t)-S_{T}(t)].
\end{eqnarray}
The Sudakov exponents are defined as
\begin{eqnarray}
S_{B}(t)\,=\,s\left(x_{1}\frac{m_{B}}{\sqrt{2}},b_{1}\right)\,+\,\frac{5}{3}\int_{1/b_{1}}^{t}\frac{d\bar{\mu}}{\bar{\mu}}
\gamma_{q}(\alpha_{s}(\bar{\mu})),
\end{eqnarray}
\begin{eqnarray}
S_{D}(t)\,=\,s\left(x_{D}\frac{m_{B}}{\sqrt{2}},b\right)
\,+\,2\int_{1/b}^{t}\frac{d\bar{\mu}}{\bar{\mu}}\gamma_{q}(\alpha_{s}(\bar{\mu})),
\end{eqnarray}
\begin{eqnarray}
S_{T}(t)\,=\,s\left(x_{T}\frac{m_{B}}{\sqrt{2}},b\right)\,+\,s\left((1-x_{T})\frac{m_{B}}{\sqrt{2}},b\right)
\,+\,2\int_{1/b}^{t}\frac{d\bar{\mu}}{\bar{\mu}}\gamma_{q}(\alpha_{s}(\bar{\mu})),
\end{eqnarray}
where the $s(Q,b)$ can be found in the Appendix A in the
Ref.\cite{prd63074009}. $x_{T(D)}$ is the momentum fraction of
``(light) quark" in tensor (D) meson.

For the rest of diagrams, the related functions are summarized as
follows:
\begin{eqnarray}
t_{c}\,&=&\,\max\{\sqrt{x_{1}x_{3}(1-r_{D}^{2})}m_{B},\sqrt{|x_{1}-x_{2}|x_{3}(1-r_{D}^{2})}m_{B},1/b_{1},1/b_{2}\},\nonumber\\
t_{d}\,&=&\,\max\{\sqrt{x_{1}x_{3}(1-r_{D}^{2})}m_{B},\sqrt{|x_{1}+x_{2}-1|(r_{D}^{2}+x_{3}(1-r_{D}^{2}))+r_{D}^{2}}m_{B},
\nonumber\\
&&1/b_{1},1/b_{2}\},
\end{eqnarray}
\begin{eqnarray}
E_{enf}(t)\,&=&\,\alpha_{s}(t)\cdot\exp[-S_{B}(t)-S_{D}(t)-S_{T}(t)]\mid\,_{b_{1}\,=\,b_{3}},
\end{eqnarray}
\begin{eqnarray}
h_{enfj}(x_{1},x_{2},x_{3},b_{1},b_{2})\,&=&\,\left[\theta(b_{2}-b_{1})K_{0}(\sqrt{x_{1}x_{3}(1-r_{D}^{2})}m_{B}b_{2})
I_{0}(\sqrt{x_{1}x_{3}(1-r_{D}^{2})}m_{B}b_{1})\right.\nonumber\\
&&\left.+\theta(b_{1}-b_{2})K_{0}(\sqrt{x_{1}x_{3}(1-r_{D}^{2})}m_{B}b_{1})I_{0}(\sqrt{x_{1}x_{3}(1-r_{D}^{2})}m_{B}b_{2})
\right]\nonumber\\
&&\cdot \left\{\begin{array}{ll}
\frac{i\pi}{2}H_{0}^{(1)}\left(\sqrt{|D_{j}^{2}|}m_{B}b_{2}\right),& \;\;D_{j}^{2}<0;\\
K_{0}\left(D_{j}m_{B}b_{2}\right),&\;\;D_{j}^{2}>0,
\end{array}\right.
\end{eqnarray}
with $j=1,2$ and $H_{0}^{(1)}(z)\,=\,J_{0}(z)\,+\,iY_{0}(z)$.
\begin{eqnarray}
D_{1}^{2}&=&(x_{1}-x_{2})x_{3}(1-r_{D}^{2})m_{B}^{2},\nonumber\\
D_{2}^{2}&=&(x_{1}+x_{2}-1)(r_{D}^{2}+x_{3}(1-r_{D}^{2}))+r_{D}^{2}.
\end{eqnarray}
\begin{eqnarray}
&&t_{e}\,=\,\max\{\sqrt{x_{3}(1-r_{D}^{2})}m_{B},1/b_{2},1/b_{3}\},\nonumber\\
&&t_{f}\,=\,\max\{\sqrt{x_{2}(1-r_{D}^{2})}m_{B},1/b_{2},1/b_{3}\},\\
&&E_{af}(t)\,=\,\alpha_{s}(t)\cdot \exp[-S_{T}(t)-S_{D}(t)],
\end{eqnarray}
\begin{eqnarray}
h_{af}(x_{2},x_{3},b_{2},b_{3})\,&=&\,(\frac{i\pi}{2})^{2}H_{0}^{(1)}\left(\sqrt{x_{2}x_{3}}m_{B}b_{2}\right)\nonumber\\
&&\left[\theta(b_{2}-b_{3})H_{0}^{(1)}\left(\sqrt{x_{3}}m_{B}b_{2}\right)J_{0}\left(\sqrt{x_{3}}m_{B}b_{3}\right)\right.
\,+\nonumber\\
&&\left.\theta(b_{3}-b_{2})H_{0}^{(1)}\left(\sqrt{x_{3}}m_{B}b_{3}\right)J_{0}\left(\sqrt{x_{3}}m_{B}b_{2}\right)\right]\cdot
S_{t}(x_{3}).
\end{eqnarray}
\begin{eqnarray}
t_{g}\,&=&\,\max\{\sqrt{x_{2}x_{3}(1-r_{D}^{2})}m_{B},\sqrt{1-(1-x_{3})(1-x_{1}-x_{2}(1-r_{D}^{2}))}m_{B},\nonumber\\
&&1/b_{1},1/b_{2}\},\nonumber\\
t_{h}\,&=&\,\max\{\sqrt{x_{2}x_{3}(1-r_{D}^{2})}m_{B},\sqrt{x_{3}|x_{1}-x_{2}(1-r_{D}^{2})|}m_{B},1/b_{1},1/b_{2}\},
\end{eqnarray}
\begin{eqnarray}
E_{anf}\,=\,\alpha_{s}(t)\cdot
\exp[-S_{B}(t)-S_{T}(t)-S_{D}(t)]\mid\,_{b_{2}=b_{3}},
\end{eqnarray}
\begin{eqnarray}
h_{anfj}(x_{1},x_{2},x_{3},b_{1},b_{2})\,&=&\,\frac{i\pi}{2}\left[\theta(b_{1}-b_{2})H_{0}^{(1)}\left(Fm_{B}b_{1}
\right)J_{0}\left(Fm_{B}b_{2}\right)\right.\nonumber\\
&&\left.+\theta(b_{2}-b_{1})H_{0}^{(1)}\left(Fm_{B}b_{2}\right)J_{0}\left(Fm_{B}b_{1}\right)\right]\nonumber\\
&&\times \left\{\begin{array}{ll}
\frac{i\pi}{2}H_{0}^{(1)}\left(\sqrt{|F_{j}^{2}|}m_{B}b_{1}\right),&
F_{j}^{2}<0,\\
K_{0}\left(F_{j}m_{B}b_{1}\right),& F_{j}^{2}>0,
\end{array}\right.
\end{eqnarray}
with $j=1,2$.
\begin{eqnarray}
F^{2}&=&x_{2}x_{3}(1-r_{D}^{2}),\nonumber\\
F_{1}^{2}&=&1-(1-x_{3})(1-x_{1}-x_{2}(1-r_{D}^{2})),\nonumber\\
F_{2}^{2}&=&x_{3}(x_{1}-x_{2}(1-r_{D}^{2})).
\end{eqnarray}
For $B\rightarrow \bar{D}^{(*)}T$ decays:
\begin{eqnarray}
t_{e(f)}^{\prime}&=&t_{e(f)}\;(x_{2}\rightarrow
1-x_{3},\;x_{3}\rightarrow x_{2}),\nonumber\\
 h_{anfj}^{\prime}&=&h_{anfj}\;(x_{3}\rightarrow
(1-x_{3})(1-r_{D}^{2}),x_{2}(1-r_{D}^{2})\rightarrow
x_{2}),\nonumber\\
t_{g(h)}^{\prime}&=&t_{g(h)}\;(x_{3}\rightarrow
(1-x_{3})(1-r_{D}^{2}),x_{2}(1-r_{D}^{2})\rightarrow x_{2}),
\end{eqnarray}
\begin{eqnarray}
t_{c}^{3}&=&\max\{\sqrt{x_{1}x_{3}}m_{B},\sqrt{|x_{1}-(1-x_{2})(1-r_{D}^{2})|x_{3}}m_{B},1/b_{1},1/b_{2}\},\nonumber\\
t_{d}^{3}&=&\max\{\sqrt{x_{1}x_{3}}m_{B},\sqrt{|x_{1}-x_{2}(1-r_{D}^{2})|x_{3}}m_{B},1/b_{1},1/b_{2}\},
\end{eqnarray}
\begin{eqnarray}
E_{enf}^{3}(x_{1},x_{2},x_{3},b_{1},b_{2})&=&\left[\theta(b_{2}-b_{1})I_{0}(\sqrt{x_{1}x_{3}}m_{B}b_{1})K_{0}(\sqrt{x_{1}x_{3}}
m_{B}b_{2})\right.\nonumber\\
&&\left.+\theta(b_{1}-b_{2})I_{0}(\sqrt{x_{1}x_{3}}m_{B}b_{2})K_{0}(\sqrt{x_{1}x_{3}}m_{B}b_{1})\right]\nonumber\\
&&\times \left\{\begin{array}{ll}
\frac{i\pi}{2}H_{0}^{(1)}\left(\sqrt{|G^{2}|}m_{B}b_{2}\right),&
G^{2}<0,\\
K_{0}\left(Gm_{B}b_{2}\right),& G^{2}>0,
\end{array}\right.
\end{eqnarray}
with $G^{2}=(x_{1}-x_{2}(1-r_{D}^{2}))x_{3}m_{B}^{2}$.

\section{ Formulas for $B\rightarrow \bar{D}^{(*)}T$ decay Amplitudes}\label{b}

 For $B \rightarrow
\bar{D} T$ decays, the expression of the factorizbale emission
contributions ($\mathcal {A}_{ef}$) is the same as Eq.(\ref{ef}).
For nonfactorizable emission     diagrams, the amplitudes are given
by
\begin{eqnarray}
\mathcal {M}_{enf}&=&-\frac{32}{3}C_{F}\pi
m_{B}^{4}\int_{0}^{1}\,dx_{1}dx_{2}dx_{3}\int_{0}^{1/\Lambda}\,b_{1}db_{1}b_{2}db_{2}\,\phi_{B}(x_{1},b_{1})
\phi_{D}(x_{2},b_{2})\nonumber\\
&&\times
\left\{\left[\phi_{T}(x_{3})(1-x_{2})+(\phi_{T}^{t}(x_{3})-\phi_{T}^{s}(x_{3}))r_{T}x_{3}\right]\right.\nonumber\\
&&\cdot \left. h_{enf2}(x_{i},b_{i})E_{enf}(t_{d})\right.\nonumber\\
&&\left.+\left[-\phi_{T}(x_{3})(x_{2}+x_{3})+(\phi_{T}^{s}(x_{3})+\phi_{T}^{t}(x_{3}))r_{T}x_{3}\right]\right.\nonumber\\
&& \cdot \left. h_{enf1}(x_{i},b_{i})E_{enf}(t_{c})\right\}.
\end{eqnarray}
The decay amplitudes of  factorizable annihilation contributions
($\mathcal {A}_{af}$) can be obtain by making the following
substitutions in Eq.(\ref{af}),
\begin{eqnarray}
x_{2}\rightarrow 1-x_{3},\;x_{3}\rightarrow
x_{2},\;\phi_{T}^{(t,s)}(x_{2})\rightarrow\phi_{T}^{(t,s)}(1-x_{3}),
\;\phi_{D}(x_{3},b_{3})\rightarrow\phi_{D}(x_{2},b_{2}).
\end{eqnarray}
While for the non-factorizable   annihilation contributions, the
decay amplitude is
\begin{eqnarray}
\mathcal {M}_{anf}&=&\frac{32}{3}C_{F}\pi
m_{B}^{4}\int_{0}^{1}\,dx_{1}dx_{2}dx_{3}\int_{0}^{1/\Lambda}\,b_{1}db_{1}b_{2}db_{2}\,\phi_{B}(x_{1},b_{1})\phi_{D}(x_{2},
b_{2})\nonumber\\
&&\times \left\{\left[\phi_{T}(x_{3})x_{2}+r_{D}r_{T}(\phi_{T}^{s}(x_{3})(x_{3}-x_{2}-3)+\phi_{T}^{t}(x_{3})(x_{2}+x_{3}-1))
\right]\right.\nonumber\\
&&\left.\cdot h_{anf1}^{\prime}(x_{1},x_{2},x_{3},b_{1},b_{2})E_{anf}(t_{g}^{\prime})\right.\nonumber\\
&&\left.+\left[\phi_{T}(x_{3})(x_{3}-1)+r_{D}r_{T}(\phi_{T}^{s}(x_{3})(x_{2}-x_{3}+1)+\phi_{T}^{t}(x_{3})(x_{2}+x_{3}-1))
\right]\right.\nonumber\\
&&\cdot\left.h_{anf2}^{\prime}(x_{1},x_{2},x_{3},b_{1},b_{2})E_{anf}(t_{h}^{\prime})\right\}.
\end{eqnarray}

For those two nonfactorizable emission diagrams in Fig.3, the decay
amplitude is
\begin{eqnarray}
\mathcal {M}_{enf}^{3}\,&=&\,\frac{32}{3}C_{F}\pi
m_{B}^{4}\int_{0}^{1}\,dx_{1}dx_{2}dx_{3}\int_{0}^{1/\Lambda}\,b_{1}db_{1}b_{2}db_{2}\,\phi_{B}(x_{1},b_{1})
\phi_{D}(x_{3},b_{1})\nonumber\\
&&\times
\left\{\left[\phi_{T}(x_{2})(x_{2}-1+r_{D}x_{3})\right]h_{enf}^{3}(x_{1},(1-x_{2}),x_{3},b_{1},b_{2})E_{enf}(t_{c}^{3})
\right.\nonumber\\
&&\left.+\left[\phi_{T}(x_{2})(x_{2}+x_{3}-r_{D}x_{3})\right]h_{enf}^{3}(x_{1},x_{2},x_{3},b_{1},b_{2})E_{enf}(t_{d}^{3})
\right\}.
\end{eqnarray}

For $B\rightarrow \bar{D}^{*} T$ decays, the expressions of the
longitudinally polarized contributions of the emission diagrams can
be obtained from
 those corresponding  $B\rightarrow \bar{D} T$
decays by substitution in Eq.\ref{tihuan}. For the annihilation type
diagrams, the decay amplitudes are
\begin{eqnarray}
\mathcal {A}_{af}^{L}&=&8\sqrt{\frac{2}{3}}\,C_{F}f_{B}\pi
m_{B}^{4}\int_{0}^{1}\,dx_{2}dx_{3}\int_{0}^{1/\Lambda}\,b_{2}db_{2}b_{3}db_{3}\,\phi_{D}^{L}(x_{2},b_{2})\nonumber\\
&&\times\left\{\left[\phi_{T}(x_{3})(1-x_{3})-r_{D}r_{T}(\phi_{T}^{s}(x_{3})+\phi_{T}^{t}(x_{3}))\right]\right.\nonumber\\
&&\left.\cdot h_{af}(1-x_{3},x_{2}(1-r_{D}^{2}),b_{2},b_{3})E_{af}(t_{e}^{\prime})\right.\nonumber\\
&&\left.+\left[-\phi_{T}(x_{3})x_{2}+2r_{D}r_{T}\phi_{T}^{s}(x_{3})(x_{2}-1)\right]\right.\nonumber\\
&&\cdot\left.h_{af}(x_{2},(1-x_{3})(1-r_{D}^{2}),b_{3},b_{2})E_{af}(t_{f}^{\prime})\right\},
\end{eqnarray}
\begin{eqnarray}
\mathcal {M}_{anf}^{L}&=&\frac{32}{3}C_{F}\pi
m_{B}^{4}\int_{0}^{1}\,dx_{1}dx_{2}dx_{3}\int_{0}^{1/\Lambda}\,b_{1}db_{1}b_{2}db_{2}\,\phi_{B}(x_{1},b_{1})
\phi_{D}^{L}(x_{2},b_{2})\nonumber\\
&&\times \left\{\left[\phi_{T}(x_{3})x_{2}-r_{D}r_{T}(\phi_{T}^{t}(x_{3})(1+x_{3}-x_{2})+\phi_{T}^{s}(x_{3})(x_{2}+x_{3}-1))
\right]\right.\nonumber\\
&&\left.\cdot h_{anf1}^{\prime}(x_{1},x_{2},x_{3},b_{1},b_{2})E_{anf}(t_{g}^{\prime})\right.\nonumber\\
&&\left.+\left[\phi_{T}(x_{3})(x_{3}-1)-r_{D}r_{T}(\phi_{T}^{t}(x_{3})(x_{2}-x_{3}+1)+\phi_{T}^{s}(x_{3})(x_{2}+x_{3}-1))
\right]\right.\nonumber\\
&&\cdot\left.h_{anf2}^{\prime}(x_{1},x_{2},x_{3},b_{1},b_{2})E_{anf}(t_{h}^{\prime})\right\}.
\end{eqnarray}
The transversely polarized contributions are   suppressed by $r_{D}$
or $r_{T}$. For the factorizable emission diagrams, the expressions
are the same as Eq.(\ref{eft}) and Eq.(\ref{efn}). For the
non-factorizable emission diagrams, the decay amplitudes are
\begin{eqnarray}
\mathcal {M}_{enf}^{T}&=&16\sqrt{\frac{1}{3}}C_{F}\pi r_{D}
m_{B}^{4}\int_{0}^{1}\,dx_{1}dx_{2}dx_{3}\int_{0}^{1/\Lambda}\,b_{1}db_{1}b_{2}db_{2}\,\phi_{B}(x_{1},b_{1})\phi_{D}^{T}
(x_{2},b_{2})\nonumber\\
&&\times
\left\{\left[\phi_{T}^{T}(x_{3})(x_{2}-1)+r_{T}(\phi_{T}^{a}(x_{3})-\phi_{T}^{v}(x_{3}))\right] h_{enf2}(x_{i},b_{i})
E_{enf}(t_{d})\right.\nonumber\\
&&\left.+\left[-\phi_{T}^{T}(x_{3})x_{2}+2r_{T}(x_{2}+x_{3})\phi_{T}^{v}(x_{3})\right]h_{enf1}(x_{i},b_{i})E_{enf}(t_{c})
\right\},
\end{eqnarray}
\begin{eqnarray}
\mathcal {M}_{enf}^{N}\,=\,\mathcal
{M}_{enf}^{T}(\phi_{T}^{a}\leftrightarrow\phi_{T}^{v}).
\end{eqnarray}
For factorizable annihilation diagrams, the transverse decay
amplitudes are
\begin{eqnarray}
\mathcal {A}_{af}^{T}&=&-4\sqrt{2}\,C_{F}f_{B}\pi
m_{B}^{4}r_{D}\int_{0}^{1}\,dx_{2}dx_{3}\int_{0}^{1/\Lambda}\,b_{2}db_{2}b_{3}db_{3}\,\phi_{D}^{T}(x_{2},b_{2})\nonumber\\
&&\times\left\{\left[r_{D}\phi_{T}^{T}(x_{3})-r_{T}(\phi_{T}^{a}(x_{3})x_{3}-\phi_{T}^{v}(x_{3})(x_{3}-2))\right]\right.
\nonumber\\
&&\left.\cdot h_{af}(1-x_{3},x_{2}(1-r_{D}^{2}),b_{2},b_{3})E_{af}(t_{e})\right.\nonumber\\
&&\left.+r_{T}\left[\phi_{T}^{v}(x_{3})(x_{2}+1)+\phi_{T}^{a}(x_{3})(x_{2}-1)\right]\right.\nonumber\\
&&\cdot\left.h_{af}(x_{2},(1-x_{3})(1-r_{D}^{2}),b_{3},b_{2})E_{af}(t_{f})\right\},
\end{eqnarray}
\begin{eqnarray}
\mathcal {A}_{af}^{N}&=&-4\sqrt{2}\,C_{F}f_{B}\pi
m_{B}^{4}r_{D}\int_{0}^{1}\,dx_{2}dx_{3}\int_{0}^{1/\Lambda}\,b_{2}db_{2}b_{3}db_{3}\,\phi_{D}^{T}(x_{2},b_{2})\nonumber\\
&&\times\left\{\left[-r_{D}\phi_{T}^{T}(x_{3})-r_{T}(\phi_{T}^{v}(x_{3})x_{3}-\phi_{T}^{a}(x_{3})(x_{3}-2))\right]\right.
\nonumber\\
&&\left.\cdot h_{af}(1-x_{3},x_{2}(1-r_{D}^{2}),b_{2},b_{3})E_{af}(t_{e})\right.\nonumber\\
&&\left.+r_{T}\left[\phi_{T}^{a}(x_{3})(x_{2}+1)+\phi_{T}^{v}(x_{3})(x_{2}-1)\right]\right.\nonumber\\
&&\cdot\left.h_{af}(x_{2},(1-x_{3})(1-r_{D}^{2}),b_{3},b_{2})E_{af}(t_{f})\right\}.
\end{eqnarray}
While for the non-factorizable annihilation diagrams, the transverse
decay amplitudes are
\begin{eqnarray}
\mathcal {M}_{anf}^{T}&=&16\sqrt{\frac{1}{3}}C_{F}\pi
m_{B}^{4}r_{D}\int_{0}^{1}\,dx_{1}dx_{2}dx_{3}\int_{0}^{1/\Lambda}\,b_{1}db_{1}b_{2}db_{2}\,\phi_{B}(x_{1},b_{1})
\phi_{D}^{T}(x_{2},b_{2})\nonumber\\
&&\times \left\{\left[r_{D}\phi_{T}^{T}(x_{3})(x_{2}-1)+2r_{T}\phi_{T}^{v}(x_{3})\right]h_{anf1}^{\prime}
(x_{1},x_{2},x_{3},b_{1},b_{2})E_{anf}(t_{g}^{\prime})\right.\nonumber\\
&&\left.-\left[r_{D}x_{2}\phi_{T}^{T}(x_{3})\right]h_{anf2}^{\prime}(x_{1},x_{2},x_{3},b_{1},b_{2})E_{anf}(t_{h}^{\prime})
\right\},
\end{eqnarray}
\begin{eqnarray}
\mathcal {M}_{anf}^{N}&=&16\sqrt{\frac{1}{3}}C_{F}\pi
m_{B}^{4}r_{D}\int_{0}^{1}\,dx_{1}dx_{2}dx_{3}\int_{0}^{1/\Lambda}\,b_{1}db_{1}b_{2}db_{2}\,\phi_{B}(x_{1},b_{1})\phi_{D}^{T}
(x_{2},b_{2})\nonumber\\
&&\times \left\{\left[-r_{D}\phi_{T}^{T}(x_{3})(x_{2}-1)+2r_{T}\phi_{T}^{a}(x_{3})\right]h_{anf1}^{\prime}(x_{1},x_{2},x_{3},
b_{1},b_{2})
E_{anf}(t_{g}^{\prime})\right.\nonumber\\
&&\left.+\left[r_{D}x_{2}\phi_{T}^{T}(x_{3})\right]h_{anf2}^{\prime}(x_{1},x_{2},x_{3},b_{1},b_{2})E_{anf}(t_{h}^{\prime})
\right\}.
\end{eqnarray}
The non-factorizable emission decay amplitudes for $B\rightarrow
\bar{D}^{*}T$ with a tensor meson emitted shown in Fig.3 are
\begin{eqnarray}
\mathcal {M}_{enf}^{3L}\,&=&\,-\frac{32}{3}C_{F}\pi
m_{B}^{4}(r_{D}-1)\int_{0}^{1}\,dx_{1}dx_{2}dx_{3}\int_{0}^{1/\Lambda}\,b_{1}db_{1}b_{2}db_{2}\,\phi_{B}(x_{1},b_{1})
\phi_{D}^{L}(x_{3},b_{1})\nonumber\\
&&\times
\left\{\left[\phi_{T}(x_{2})(x_{2}-1+r_{D}(x_{2}-x_{3}-1))\right]h_{enf}^{3}(x_{1},(1-x_{2}),x_{3},b_{1},b_{2})E_{enf}
(t_{c}^{3})\right.\nonumber\\
&&\left.+\left[\phi_{T}(x_{2})(x_{2}+x_{3}+r_{D}x_{2})\right]h_{enf}^{3}(x_{1},x_{2},x_{3},b_{1},b_{2})E_{enf}(t_{d}^{3})
\right\},
\end{eqnarray}
\begin{eqnarray}
\mathcal {M}_{enf}^{3T}\,&=&\,16\sqrt{\frac{1}{3}}C_{F}\pi
m_{B}^{4}r_{T}\int_{0}^{1}\,dx_{1}dx_{2}dx_{3}\int_{0}^{1/\Lambda}\,b_{1}db_{1}b_{2}db_{2}\,\phi_{B}(x_{1},b_{1})
\phi_{D}^{T}(x_{3},b_{1})\nonumber\\
&&\times
\left\{\left[(\phi_{T}^{a}(x_{2})+\phi_{T}^{v}(x_{2}))(x_{2}-1)\right]h_{enf}^{3}(x_{1},(1-x_{2}),x_{3},b_{1},b_{2})
E_{enf}(t_{c}^{3})\right.\nonumber\\
&&\left.+\left[-\phi_{T}^{a}(x_{2})x_{2}+\phi_{T}^{v}(x_{2})((2r_{D}-1)x_{2}+2r_{D}x_{3})\right]\right.\nonumber\\
&&\left.\cdot
h_{enf}^{3}(x_{1},x_{2},x_{3},b_{1},b_{2})E_{enf}(t_{d}^{3})\right\},
\end{eqnarray}
\begin{eqnarray}
\mathcal {M}_{enf}^{3N}=-\mathcal
{M}_{enf}^{3T}(\phi_{T}^{a}\leftrightarrow \phi_{T}^{v}).
\end{eqnarray}
With the functions obtained in the above, the amplitudes of all
$B\rightarrow \bar{D}T$ decay channels can be given by
\begin{eqnarray}
&&\mathcal {A}(B^{0}\rightarrow
\bar{D}^{0}a_{2}^{0})=\frac{G_{F}}{\sqrt{2}}\frac{1}{\sqrt{2}}V_{cb}^{*}V_{ud}[a_{2}\mathcal
{A}_{af}+C_{2}\mathcal {M}_{anf}-a_{2}\mathcal
{A}_{ef}-C_{2}\mathcal {M}_{enf}],\\
&&\mathcal {A}(B^{0}\rightarrow
\bar{D}^{0}f_{2}^{q})=\frac{G_{F}}{\sqrt{2}}\frac{1}{\sqrt{2}}V_{cb}^{*}V_{ud}[a_{2}\mathcal
{A}_{af}+C_{2}\mathcal {M}_{anf}+a_{2}\mathcal
{A}_{ef}+C_{2}\mathcal {M}_{enf}],\\
&&\mathcal {A}(B^{0}\rightarrow
\bar{D}^{0}K_{2}^{*0})=\frac{G_{F}}{\sqrt{2}}V_{cb}^{*}V_{us}[a_{2}\mathcal
{A}_{ef}+C_{2}\mathcal {M}_{enf}],\\
&&\mathcal {A}(B^{0}\rightarrow
D^{-}a_{2}^{+})=\frac{G_{F}}{\sqrt{2}}V_{cb}^{*}V_{ud}[C_{1}\mathcal
{M}_{enf}^{3}+a_{2}\mathcal
{A}_{af}+C_{2}\mathcal {M}_{anf}],\\
&&\mathcal {A}(B^{0}\rightarrow
D^{-}K_{2}^{*+})=\frac{G_{F}}{\sqrt{2}}V_{cb}^{*}V_{us}[C_{1}\mathcal
{M}_{enf}^{3}],\\
&&\mathcal {A}(B^{0}\rightarrow
D_{s}^{-}K_{2}^{*+})=\frac{G_{F}}{\sqrt{2}}V_{cb}^{*}V_{ud}[a_{2}\mathcal
{A}_{af}+C_{2}\mathcal {M}_{anf}],\\
&&\mathcal {A}(B^{+}\rightarrow
\bar{D}^{0}a_{2}^{+})=\frac{G_{F}}{\sqrt{2}}V_{cb}^{*}V_{ud}[a_{2}\mathcal
{A}_{ef}+C_{2}\mathcal {M}_{enf}+C_{1}\mathcal {M}_{enf}^{3}],\\
&&\mathcal {A}(B^{+}\rightarrow
\bar{D}^{0}K_{2}^{*+})=\frac{G_{F}}{\sqrt{2}}V_{cb}^{*}V_{us}[a_{2}\mathcal
{A}_{ef}+C_{2}\mathcal {M}_{enf}+C_{1}\mathcal {M}_{enf}^{3}],\\
&&\mathcal {A}(B_{s}^{0}\rightarrow
\bar{D}^{0}f_{2}^{q})=\frac{G_{F}}{\sqrt{2}}\frac{1}{\sqrt{2}}V_{cb}^{*}V_{us}[a_{2}\mathcal
{A}_{af}+C_{2}\mathcal {M}_{anf}],\\
&&\mathcal {A}(B_{s}^{0}\rightarrow
\bar{D}^{0}a_{2}^{0})=\frac{G_{F}}{\sqrt{2}}\frac{1}{\sqrt{2}}V_{cb}^{*}V_{us}[a_{2}\mathcal
{A}_{af}+C_{2}\mathcal {M}_{anf}],
\end{eqnarray}
\begin{eqnarray}
&&\mathcal {A}(B_{s}^{0}\rightarrow
\bar{D}^{0}f_{2}^{s})=\frac{G_{F}}{\sqrt{2}}V_{cb}^{*}V_{us}[a_{2}\mathcal
{A}_{ef}+C_{2}\mathcal {M}_{enf}],\\
&&\mathcal {A}(B_{s}^{0}\rightarrow
\bar{D}^{0}\bar{K}_{2}^{*0})=\frac{G_{F}}{\sqrt{2}}V_{cb}^{*}V_{ud}[a_{2}\mathcal
{A}_{ef}+C_{2}\mathcal {M}_{enf}],\\
&&\mathcal {A}(B_{s}^{0}\rightarrow
D^{-}a_{2}^{+})=\frac{G_{F}}{\sqrt{2}}V_{cb}^{*}V_{us}[a_{2}\mathcal
{A}_{af}+C_{2}\mathcal {M}_{anf}],\\
&&\mathcal {A}(B_{s}^{0}\rightarrow
D_{s}^{-}a_{2}^{+})=\frac{G_{F}}{\sqrt{2}}V_{cb}^{*}V_{ud}[C_{1}\mathcal {M}_{enf}^{3}],\\
&&\mathcal {A}(B_{s}^{0}\rightarrow
D_{s}^{-}K_{2}^{*+})=\frac{G_{F}}{\sqrt{2}}V_{cb}^{*}V_{us}[a_{2}\mathcal
{A}_{af}+C_{2}\mathcal {M}_{anf}+C_{1}\mathcal {M}_{enf}^{3}].
\end{eqnarray}

\end{appendix}


\begin{thebibliography}{99}
\bibitem{jpg37075021}
K. Nakamura et al. [Particle Data Group], J Phys. G \textbf{37},
075021 (2010).
\bibitem{wwprd83014008}
Wei Wang, Phys. Rev. D \textbf{83}, 014008 (2011).
\bibitem{epjc22683}
C. S. Kim, B. H. Lim and S. Oh, Eur. Phys. J. C \textbf{22}, 683
(2002) [arXiv:hep-ph/0101292].
\bibitem{epjc22695}
C. S. Kim, B. H. Lim and S. Oh, Eur. Phys. J. C \textbf{22}, 695
(2002) [Erratum-ibid. C \textbf{24}, 665 (2002)]
[arXiv:hep-ph/01080504].
\bibitem{prd67014002}
C. S. Kim, B. H. Lim and S. Oh, Phys. Rev. D \textbf{67}, 014002
(2003) [arXiv:hep-ph/0205263].
\bibitem{jpg36095004}
J. H. Mu\`{n}oz and N. Quintero, J. Phys. G \textbf{36}, 095004
(2009) [arXiv:0903.3701 [hep-ph]].
\bibitem{arxiv1004.1928}
N. Sharma and R. C. Verma, arXiv:1004.1928 [hep-ph].
\bibitem{zheng1}
H. Y. Cheng, Y. Koike and K. C. Yang, Phys. Rev. D \textbf{82},
054019 (2010) [arXiv:1007.3541 [hep-ph]].
\bibitem{zheng2}
Hai-Yang Cheng and Kwei-Chou Yang, Phys. Rev. D \textbf{83}, 034001
(2011) [arXiv:1010.3309 [hep-ph]].
\bibitem{zou}
Z. T. Zou, X. Yu and C. D. L\"{u}, arXiv:1203.4120 [hep-ph].

\bibitem{prd491645}
A. C. Katoch and R. C. Verma, Phys. Rev. D \textbf{49}, 1645
(1994); \textbf{52}, 1717 (1995); \textbf{55}, 7315(E) (1997).
\bibitem{prd555581}
G. L\'{o}pez Castro and J. H. Mu\~{n}oz, Phys. Rev. D \textbf{55},
5581 (1997) [arXiv:hep-ph/9702238].
\bibitem{prd59077504}
J. H. Mu\~{n}oz, A. A. Rojas, and G. L\'{o}pez Castro, Phys. Rev. D
\textbf{59}, 077504 (1999).

\bibitem{arxiv1010.3077}
N. Sharma, R. Dhir and R. C. Verma, Phys. Rev. D \textbf{83}, 014007
(2011) [arXiv:1010.3077 [hep-ph]].
\bibitem{prd67014011}
C. S. Kim, B. H. Lim and S. Oh, Phys. Rev. D \textbf{67}, 014011
(2003).
\bibitem{prd85051301}
Wei Wang, Phys. Rev. D \textbf{85}, 051301 (2012).
\bibitem{zou-zhou}
Z. T. Zou, Zhou Rui and C. D. L\"{u}, arXiv:1204.3144 [hep-ph].
\bibitem{wang7}
Y. Y. Keum, H. n. Li and A. I. Sanda, Phys. Lett. B \textbf{504}, 6
(2001) [arXiv:hep-ph/0004004]; Phys. Rev. D \textbf{63}, 054008
(2001) [arXiv:hep-ph/0004173].
\bibitem{prd63074009}
 C. D. L\"{u}, K. Ukai and M. Z. Yang, Phys. Rev. D \textbf{63},
074009 (2001) [arXiv:hep-ph/0004213].


\bibitem{annihilation1}
C. D. L\"{u}, K. ukai, Eur. Phys. J. C \textbf{28}, 305 (2003)
[arXiv:hep-ph/0210206].
\bibitem{annihilation2}
Y. Li, C. D. L\"{u}, J. Phys. G \textbf{29}, 2115 (2003)
[arXiv:hep-ph/0304288]; High Energy Phys. \& Nucl. Phys.
\textbf{27}, 1062 (2003) [arXiv:hep-ph/0305278].
\bibitem{fd1}
H. n. Li, Phys. Rev. D \textbf{52}, 3958 (1995)
[arXiv:hep-ph/9412340]; C. Y. Wu, T. W. Yeh and H. n. Li, Phys. Rev.
D \textbf{53}, 4982 (1996) [arXiv:hep-ph/9510313]; Y. Y. Keum et
al., Phys. Rev. D \textbf{69}, 094018 (2004); C. D. L\"{u}, Phys.
Rev. D \textbf{68}, 097502 (2003).
\bibitem{fd2}
C. D. L\"{u}, Eur. Phys. J. C \textbf{24}, 121-126 (2002); J. F.
Cheng, D. S. Du and C. D. L\"{u}, Eur. Phys. J. C \textbf{45}, 711
(2006).
\bibitem{scet}
C. W. Bauer, D. Pirjol, I. Z. Rothstein and I. W. Stewart, Phys. Rev. D \textbf{70}, 054015 (2004).
\bibitem{rmp68}
G. Buchalla, A. J. Buras and M. E. Lautenbacher, \textbf{ Rev. Mod.
Phys. 68} (1996) 1125; A.J. Buras, [hep-ph/9806471].


\bibitem{prd66094010}
H. N. Li, Phys. Rev. D \textbf{66}, 094010 (2002)
\bibitem{prd57443}
H. N. Li and B. Tseng, Phys. Rev. D \textbf{57}, 443 (1998)
\bibitem{lvepjc23275}
C. D. L\"{u} and M. Z. Yang, Eur. Phys. J. C \textbf{23}, 275-287
(2002).

\bibitem{epjc28515}
C. D. L\"{u} and M. Z. Yang, Eur. Phys. J. C \textbf{28}, 515
(2003).
\bibitem{prd65014007}
T. Kurimoto, H. n. Li and A. I. Sanda, Phys. Rev. D \textbf{65},
014007 (2002).
\bibitem{wang13}
Z. T. Wei and M. Z. Yang; Nucl. Phys. B \textbf{642}, 263 (2002).

\bibitem{wang14}
A. Ali, et al., Phys. Rev. D \textbf{76}, 074018 (2007).
\bibitem{prd67054028}
T. Kurimoto, H. n. Li and A. I. Sanda, Phys. Rev. D \textbf{67}
054028 (2003).
\bibitem{liprd78014018}
R. H. Li, C. D. L\"{u} and H. Zou, Phys. Rev. D \textbf{78} 014018
(2008).
\bibitem{zouhaojpg37}
H. Zou, R. H. Li, X. X. Wang and C. D. L\"{u}, J. Phys. G \textbf{37},
015002 (2010).
\bibitem{liprd81034006}
R. H. Li, C. D. L\"{u}, A. I. Sanda, X. X. Wang, Phys. Rev. D
\textbf{81}, 034006 (2010).
\bibitem{fd}
E. Follana, C. T. H. Davies, G. P.Lepage and J. Shigemitsu [HPQCD
Collaboration and UKQCD Collaboration], Phys. Rev. Lett.
\textbf{100}, 062002 (2008).
\bibitem{am}
A. V. Manohar and M. B. Wise, Camb. Monogr. Part. Phys. Nucl. Phys.
Cosmol. 10, 1 (2000).
\bibitem{helicity1}
J. G. K\"{o}rner and G. R. Goldstein, Phys. Lett. \textbf{B79}, 105 (1979).
\bibitem{helicity2}
A. L. Kagan, Phys. Lett. \textbf{B601}, 151 (2004).
\bibitem{polarization}
A. V. Gritsan, eConf. C \textbf{070512},001 (2007).
\bibitem{TM}
T. M. Aliev and M. A. Shifman, Phys. Lett. B \textbf{112}, 401
(1982); Sov. J. Nucl. Phys. \textbf{36}, 981 (1982) [Yad. Fiz.
\textbf{36}, 1532 (1982)].
\bibitem{TK}
T. M. Aliev, K. Azizi and V. Bashiry, J. Phys. G \textbf{37}, 025001
(2010).
 \textbf{48}, 339 (1993).
 \bibitem{zheng3}
 Hai-Yang Cheng and Robert Shrock, Phys. Rev. D \textbf{84}, 094008
 (2011) [arXiv:hep-ph/1109.3877].
\bibitem{jpg27807}
D. M. Li, H. Yu, and Q. X. Shen, J. Phys. G \textbf{27}, 807 (2001).
\bibitem{plb555}
H.-n. Li and K. Ukai, Phys. Lett. B \textbf{555}, 197 (2003).



\end{thebibliography}
\end{document}